\documentclass{aa}
\usepackage{graphicx}
\usepackage{bm,natbib,url,wasysym}
\usepackage[varg]{txfonts}
\usepackage{xcolor}
\usepackage{amsmath}
\usepackage[colorlinks,allcolors=blue]{hyperref}
\bibpunct{(}{)}{;}{a}{}{,} % to follow the A&A style
\graphicspath{{./figs/}{./png/}}
\newcommand{\brac}[1]{\langle #1 \rangle}
\newcommand{\EQ}{\begin{equation}}
\newcommand{\EN}{\end{equation}}
\newcommand{\EQA}{\begin{eqnarray}}
\newcommand{\ENA}{\end{eqnarray}}

\newcommand{\Eq}[1]{Equation~(\ref{#1})}
\newcommand{\Eqs}[2]{Equations~(\ref{#1}) and~(\ref{#2})}
\newcommand{\Eqss}[2]{Equations~(\ref{#1})--(\ref{#2})}

\newcommand{\App}[1]{Appendix~\ref{#1}}
\newcommand{\Sec}[1]{Section~\ref{#1}}

\newcommand{\Fig}[1]{Fig.~\ref{#1}}

\newcommand{\Figs}[2]{Figs.~\ref{#1} and \ref{#2}}

\newcommand{\Tab}[1]{Table~\ref{#1}}

\newcommand{\bra}[1]{\langle #1\rangle}
\newcommand{\bbra}[1]{\left\langle #1\right\rangle}
\newcommand{\fluc}[1]{#1^\prime}

\newcommand{\mean}[1]{\overline#1}
\newcommand{\meanrho}{\overline{\rho}}

{}
{}
{}

\newcommand{\meanuu}{\overline{\mbox{\boldmath $u$}}{}}{}
{}
{}
{}
{}
{}
{}
{}
{}
\newcommand{\meanBB}{\overline{\mbox{\boldmath $B$}}{}}{}
{}
{}
{}
{}
{}
{}
{}
{}

{}
{}
{}

\newcommand{\meanB}{\overline{B}}

\newcommand{\meanF}{\overline{F}}

\newcommand{\meanU}{\overline{U}}

{}
{}
{}
{}

{}

{}
{}

\newcommand{\Ot}{\tilde{\Omega}}

\newcommand{\eee}{\hat{\mbox{\boldmath $e$}} {}}

\newcommand{\uu}{\mbox{\boldmath $u$} {}}

\newcommand{\bb}{\mbox{\boldmath $b$} {}}
\newcommand{\BB}{\mbox{\boldmath $B$} {}}

\newcommand{\JJ}{\mbox{\boldmath $J$} {}}

\newcommand{\AAA}{\mbox{\boldmath $A$} {}}

\newcommand{\ff}{\mbox{\boldmath $f$} {}}

\newcommand{\FF}{\mbox{\boldmath $F$} {}}

\newcommand{\FFF}{\mbox{\boldmath ${\cal F}$} {}}

\newcommand{\nab}{\mbox{\boldmath $\nabla$} {}}
\newcommand{\OO}{\bm{\Omega}}
\newcommand{\oo}{\mbox{\boldmath $\omega$} {}}

\newcommand{\SSSS}{\mbox{\boldmath ${\sf S}$} {}}

\newcommand{\sgn}{{\rm sgn}  \, {}}
\newcommand{\DD}{{\rm D} {}}

\newcommand{\dd}{{\rm d} {}}

\def\degr{\hbox{$^\circ$}}

\def\Ta{\mbox{\rm Ta}}
\def\Ra{\mbox{\rm Ra}}
\def\Ma{\mbox{\rm Ma}}
\def\Co{\mbox{\rm Co}}

\def\PrSGS{\mbox{\rm Pr}_{\rm SGS}}
\def\Pr{\mbox{\rm Pr}}

\def\Pm{\mbox{\rm Pr}_{\rm M}}
\def\Rm{\mbox{\rm Re}_{\rm M}}

\def\Rey{\mbox{\rm Re}}

\def\Co{\mbox{\rm Co}}

\def\cs{c_{\rm s}}

\def\kf{k_{\rm f}}

\def\urms{u_{\rm rms}}
\def\urmsp{u^\prime_{\rm rms}}
\def\brmsp{B^\prime_{\rm rms}}

\def\Beq{B_{\rm eq}}

\def\half{{\textstyle{1\over2}}}

\newcommand{\s}{\,{\rm s}}

\newcommand{\m}{\,{\rm m}}

\newcommand{\kg}{\,{\rm kg}}

\newcommand{\chiS}{\chi_{\rm SGS}}
\newcommand{\chiSm}{\chi_{\rm m}^{\rm SGS}}

\newcommand{\pho}{\phantom{1}}
\newcommand{\rkm}{r_{k_-}}
\newcommand{\rkp}{r_{k_+}}
\newcommand{\tkm}{\theta_{k_-}}
\newcommand{\tkp}{\theta_{k_+}}
\newcommand{\pkm}{\phi_{k_-}}
\newcommand{\pkp}{\phi_{k_+}}
\begin{document}

\titlerunning{SSD and LSD in global convection simulations}
\authorrunning{Warnecke et al.}

\title{Small-scale and large-scale dynamos in global convection
  simulations of solar-like stars}
\author{J. Warnecke\inst{1,2},  M. J.\ Korpi-Lagg\inst{1,2,3},
  M. Rheinhardt\inst{1} , M. Viviani\inst{2,4},\and A. Prabhu\inst{2,5}}
\institute{Department of Computer Science, Aalto University,
PO Box 15400, FI-00\ 076 Espoo, Finland\label{inst1}\\
\email{jorn.warnecke@aalto.fi}
\and Max-Planck-Institut für Sonnensystemforschung,
  Justus-von-Liebig-Weg 3, D-37077 G\"ottingen, Germany\\
\email{warnecke@mps.mpg.de}\label{inst2}
\and Nordita, KTH Royal Institute of Technology \& Stockholm University,
Hannes Alfv\'ens v\"ag 12, SE-11419 Stockholm, Sweden\label{inst3}
\and Wish s.r.l., Via Venezia 24, 87036 Rende (CS), Italy\label{inst4}
\and Sensirion Connected Solutions, Laubisrütistrasse 50, 8712 Stäfa, Switzerland\label{inst5}
}

\date{\today,~ $ $Revision: 1.190 $ $}
%\date{Received 12 June 2024 / Accepted ????}
\abstract{
It has been recently shown numerically that a small-scale dynamo (SSD)
instability could be possible in solar-like low magnetic Prandtl
number plasmas. It has been proposed that the presence of SSD
can potentially have a significant impact on the dynamics of
the large-scale dynamo (LSD) in the stellar convection zones. Studying
these two dynamos, SSD and LSD, together in a global magnetoconvection
model requires high-resolution simulations and large amounts of
computational resources.
}{
Starting from a well-studied global convective dynamo model that
produces cyclic magnetic fields, we systematically increased the
resolution and lowered the diffusivities to enter the regime of
Reynolds numbers that enable
the excitation of SSD on top of the
LSD. We studied how the properties of convection, generated differential
rotation profiles, and LSD solutions change with the presence of SSD.
}{
We performed semi-global convective dynamo simulations in a spherical
wedge with the Pencil Code. The resolutions of the models were
increased in 4 steps by a  total factor of 16 to achieve maximal fluid and
magnetic Reynolds numbers of above 500.
}{
We found that the differential rotation is strongly quenched by the
presence of the LSD and SSD.
Even though the small-scale magnetic
field only mildly decreases with increasing  Reynolds number, the large-scale field
strength decreases significantly.
We do not find the SSD dynamo
significantly quenching the convective flows as claimed recently by
other authors; in contrast, the convective flows first grow and then
saturate for increasing Reynolds number. Furthermore, the angular momentum
transport is highly affected by the presence of small-scale magnetic
fields, which are mostly generated by LSD. These fields not only change the Reynolds stresses, but also
generate dynamically important Maxwell stresses.
The LSD evolution in terms of its pattern and field distribution
is rather independent of the increase in
the fluid and magnetic Reynolds numbers.
}{
At high fluid and magnetic Reynolds numbers, an SSD, in addition to
the LSD, can be excited, and both have a strong influence on angular
momentum transport. Hence, it is important to study both dynamos and
their interplay together to fully understand the dynamics of the Sun
and other stars.
 }

\keywords{Magnetohydrodynamics (MHD) -- turbulence -- dynamo -- Sun:
 magnetic fields -- Stars: magnetic fields -- Stars: activity
}

\maketitle

\section{Introduction}

The Sun and other cool stars exhibit large-scale magnetic fields,
which in some cases show cyclic variations
\citep[e.g.][]{Baliunas1995, BMJRCMPW17, OLKPG17}. This is
associated with a large-scale dynamo (LSD) operating in the stellar
convection zones producing complex magnetic surface features
\citep[][]{Ch14}.
Besides the LSD, a small-scale dynamo (SSD)
has been proposed to be present in
these stars \citep[e.g.][]{RempelSSR23}.
In contrast to an LSD, the SSD does not need any
large-scale rotation, shear or stratification to operate and the
scales of its magnetic field are at or below the
characteristic scales of the flow \citep{BS05}.
There is some debate whether or not an SSD can operate in solar-like
stars.
These doubts were supported on one hand by numerical simulations,
which show that an SSD is increasingly harder to excite when approaching
the solar parameters \citep{SHBCMM05}, and on the other hand by the
inconclusive results of small-scale field observations on the Sun
\citep{BRLOSD19}.
Some observational studies show a cyclic modulation of the
inter-network field, hence a connection to the cyclic large-scale
magnetic field, whereas some studies show that the these fields are
rather cyclic independent.
The doubts based on numerical simulations, however, were recently
alleviated
by high-resolution simulations at magnetic Prandtl numbers approaching
the solar value closer than ever before \citep{WKGR23}. These simulations show that an
SSD is not only possible at these very low magnetic Prandtl numbers,
but becomes even easier to excite in this limit,
pointing to a possible
dynamical importance of the SSD in the Sun and solar-like stars.

The influence of SSD on the dynamics and LSD has been studied
with simplified setups \citep{VC92,TC13,SB15,SRB17,VPKRSK21},
while only recently  global convective dynamo simulations
\citep{Review:2023} were able to reach the regime
where both LSD and SSD are excited and can therefore
be studied
together \citep[hereafter HK21, HKS22]{HRY16,KKOWB17,HK21,HKS22}.
In \cite{HRY16}, the
authors found that the SSD suppresses
small-scale flows mimicking properties of an enhanced magnetic
diffusivity, this in turn enhances the LSD.
\cite{KKOWB17}, on the other hand found that the SSD quenches
differential rotation, which being one of the main dynamo drivers
consequently suppressed the LSD in some of their cases.

The influence of the magnetic field on the differential rotation
has been investigated in many previous analytical and numerical studies.
Already in the early global magnetoconvection studies of \cite{G83}, it was
noted that large-scale magnetism quenched differential rotation, while 
not affecting the convective motions in a clear manner. Analytical studies
in the mean-field ($\Lambda$ effect) framework produced similar
results \citep[e.g.][]{KRK1994}. Many other global magnetoconvection
studies followed
\citep[e.g.][]{FF14,KKKBOP2015,KKOBWKP16,WKMB13,WKKB16,
  BSNPVACT22,Review:2023},
yet the most relevant for the present work is \cite{KKOWB17}, because of
the presence of an SSD.
In this study the authors
found that an increase in $\Rm$ leads to strong suppression
of the differential rotation. The authors
suggested further that the Maxwell stresses become comparable to the Reynolds stresses
and therefore only a weak differential rotation can be generated.
They found that the increase in Maxwell stresses is partly due to a
strong SSD being present.
HK21 and HKS22 found that the efficient SSD in their simulation is
able to reshape an anti-solar differential rotation into a solar one
for increasing Reynolds numbers.
Taken that the results obtained with different approaches diverge quite significantly,
it is necessary to further investigate the role of SSD in LSD-active and
differentially rotating systems. As magnetic fluctuations originate
from different sources, namely SSD-action itself and tangling
of the mean field through convective turbulence, it is also important
to gain further insights into the role of these two contributions
separately. These are among the most important goals of this paper.

Since the pioneering work of \cite{B16} and \cite{KRBAKLOW17}, it
has been known
that a more realistic description of the radiative heat
diffusivity using a Kramers' opacity based term
can lead to the formation of sub-adiabatic layers at the base of the
convection zone  \citep[e.g.][]{KVKBS19,VK21}.
How the shape and depth of these layers depend on the Reynolds number or
the presence of SSD and LSD has only been studied in a few Cartesian
simulations \citep{H2017,K2019,K2021} but not in a global setup.

In this paper we present a study of a global convective dynamo model
in an azimuthal wedge of a spherical shell
using a Kramers' opacity based heat conductivity.
We increased the resolution systematically from 128x256x128 to
2048x4096x2048 grid points to reach fluid and magnetic Reynolds
numbers ($\Rm$) of above 500. This parameter regime allows us to
investigate the LSD and SSD interaction in detail.
To separately study the effect of the LSD and SSD on the overall
dynamics, we run
for each configuration: a full model, where potentially
both LSD and SSD can be excited; a "reduced"
model where the large-scale field is taken out to 
allow only for an SSD;
a hydrodynamic model in which no magnetic field is present.
The paper is organized as follows:
in Sect.~\ref{sec:model} we present our model and setup, in
Sect.~\ref{sec:results} we discuss in detail the results, and in
Sect.~\ref{sec:conclusions} we present our conclusions.
Additional information is given in Apps.~\ref{sec:diffprof}
to \ref{sec:ban}.

\begin{table*}[t!]\caption{
Summary of runs.
}\vspace{0pt}\centerline{\begin{tabular}{lcrrrrcrr|crrc}
Run & Resolution&$\Ta$[$10^{8}$]&  $\mean{\Ra}_{\rm Kram}$[$10^{7}$]& $\mean{\Ra}_{\rm SGS}$[$10^{7}$]&$\widetilde{\Ra}_{\rm SGS}$&$\Pr$&$\Rey$&$\Co$\phantom{.}&SLD&$\Delta{\theta}_\nu$&$\Delta{\theta}_\eta$&$\Delta_{\nu\eta}$\\[.8mm]   \hline
\hline\\[-2.5mm]
0M         &  $\pho128\times\pho256\times\pho128$& 1.25 & 73& 1.2 &48&5.13&  27 & 10.4&& \\
0H      &  $\pho128\times\pho256\times\pho128$& 1.25 &  73&1.2 &48&5.13&  28 & 9.9 &yes\\
\hline
1M         &  $\pho256\times\pho512\times\pho256$& 5.00 &  280 & 5.9 &94&1.52&  61 & 9.3 &&$5^\circ$&$5^\circ$&2\\
1H      &  $\pho256\times\pho512\times\pho256$& 5.00 &   280& 5.9 &94&1.52&  66 & 8.5 &yes\\
\hline
2M         &  $\pho512\times1024\times\pho512$&  20.0 &  1050 & 23.7 &149&0.58& 127 & 8.9 &&$5^\circ$&$5^\circ$&5\\
2S   &  $\pho512\times1024\times\pho512$&  20.0 &  1050  & 23.7 &149&0.58& 139 & 8.1 &&$5^\circ$&$5^\circ$&5\\
2H      &  $\pho512\times1024\times\pho512$&  20.0 &   1050  &23.7 &149&0.58& 143 & 7.9 &yes&$5^\circ$&&5\\
\hline
3M         &  $1024\times2048\times1024$&  80.0 & 3155 & 71.5 &178&0.27 &  255 & 8.9&&$17^\circ$&$17^\circ$&5\\
3S          &  $1024\times2048\times1024$&  80.0 & 3155 & 71.5 &178&0.27 &  265 & 8.5&&$17^\circ$&$5^\circ$&5\\
3H          &  $1024\times2048\times1024$&  80.0 & 3155 & 71.5 &178&0.27&  287 & 7.9&yes &$17^\circ$&&5\\
\hline
4M           &  $2048\times4096\times2048$&320.0 &7536 & 185.9&184&0.12& 517& 8.7&&$17^\circ$&$5^\circ$&5 \\
4M2         &  $2048\times4096\times2048$&320.0 &7536 & 185.9&184&0.12& 549& 8.5&&$17^\circ$&$5^\circ$&5 \\
4S           &  $2048\times4096\times2048$&320.0  &7536 & 185.9&184&0.12& 550& 8.2&&$17^\circ$&$5^\circ$&5\\
\hline
\hline
\label{runs}\end{tabular}}\tablefoot{
Columns 2 to 7:
input parameters. Columns 8 and 9:
solution parameters, calculated from the saturated
stage of the simulations. The last four columns indicate whether
slope-limited diffusion (SLD) and/or a diffusion profile in $\theta$ was
used including the corresponding parameters,
see Appendices~\ref{sec:diffprof} and \ref{sec:sld}.
All runs have a density contrast
$\Gamma_{\rho}\equiv\langle\rho\rangle_{\theta\phi}(0.7R)/\langle\rho\rangle_{\theta\phi}(R)$
of roughly 30, and $\Ot=5$, $\PrSGS=1$.
The M and S runs have $\Pm=1$.
}
\end{table*}

\section{Model and setup}
\label{sec:model}

The stellar convection zone is modeled in spherical
geometry ($r,\theta,\phi$) as a shell with a depth 
$D$ similar to that in the Sun ($0.7R \le r \le R$)
where $R$ is the radius of the star. We leave out the poles
($\theta_0\le\theta\le\pi-\theta_0$ with $\theta_0=15^\circ$) and
restrict our model to a quarter of the sphere
(a ``wedge", with
$0\le\phi\le\pi/2$),
both for numerical reasons.
Our model is similar to the ones of
\cite{KMCWB13}, \cite{KKOBWKP16}, and \cite{KVKBS19}
and we refer to these papers for a detailed description.

We solve the fully compressible MHD equations in terms of the vector potential
$\AAA$ (ensuring the solenoidality of $\BB$), the velocity $\uu$, the
specific
entropy $s$, and the density $\rho$, and employ an ideal-gas equation of state.
We include the rotational effects by adding the Coriolis force
$2\uu\times\OO_0$ to the momentum equation where
$\OO_0=\Omega_0(\cos\theta,-\sin\theta,0)$ and $\Omega_0$ is the
angular velocity of the modeled
star's co-rotating
frame, in which the plasma has zero
total angular momentum.
We choose constant magnetic diffusivity $\eta$ and
kinematic viscosity $\nu$,
except near the latitudinal boundaries, where we add in some of the runs
for numerical reasons
$\nu$ and $\eta$ profiles,
increasing with $\theta$
towards the boundaries
across an interval $\Delta\theta$,
see \App{sec:diffprof} for
details.
In our model, the diffusive heat flux has two contributions.
The first one models the radiative heat flux as $\FF^{\rm rad}=-K\nab T$
with temperature and density dependent radiative heat conductivity $K$,
based on Kramers' opacity, so that $K$ is given by
\begin{equation}
K(\rho,T)=K_0 \left({\rho\over\rho_0}\right)^{-2} \left({T\over T_0}\right)^{13/2},
\label{eq:Kkram}
\end{equation}
and the reference values $\rho_0$ and $T_0$ are set to the corresponding
values at the bottom of the domain in the initial (hydrostatic) state; for details see
\cite{BB13}, \cite{KVKBS19} and \cite{VK21}.
The second contribution mimics the heat flux of the unresolved
or sub-grid-scale (SGS) convection and stabilizes the system. This SGS heat
flux is given by $\FF^{\rm SGS}= -\chiS\rho T \nab s$.
As in \cite{KMCWB13}, $\chiS$ follows a smooth radial profile, which is zero
at the bottom of the domain, constant ($\chiS=\chiSm$) in the bulk
and maximal near the top transporting there
the majority of the heat.
For some of the hydrodynamic runs, we needed to add
a slope-limited
diffusion, acting on density and velocity, to stabilize the system, see
\App{sec:sld} for details.

\begin{figure*}[t!]
  \begin{center}
 \includegraphics[width=0.32\textwidth]{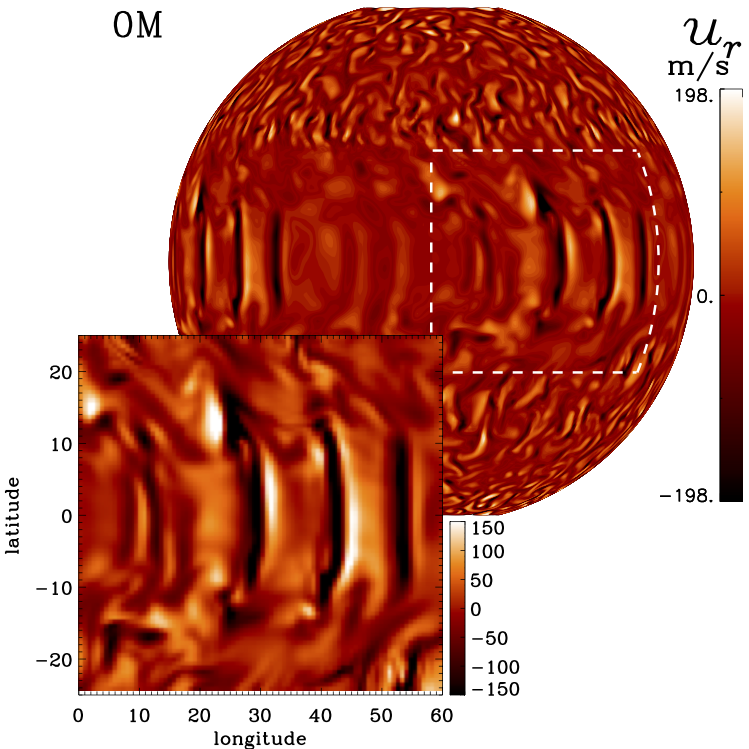}
 \includegraphics[width=0.32\textwidth]{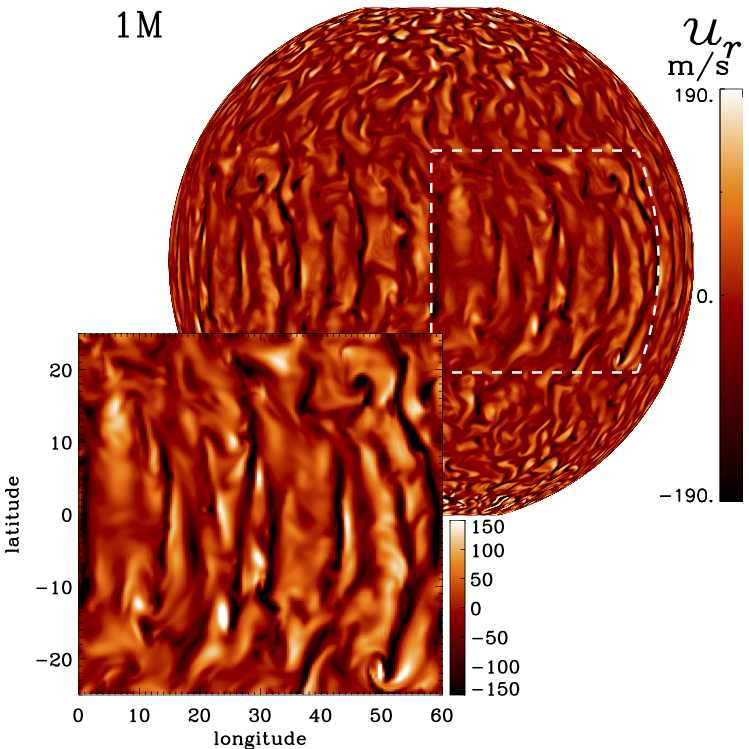}
 \includegraphics[width=0.32\textwidth]{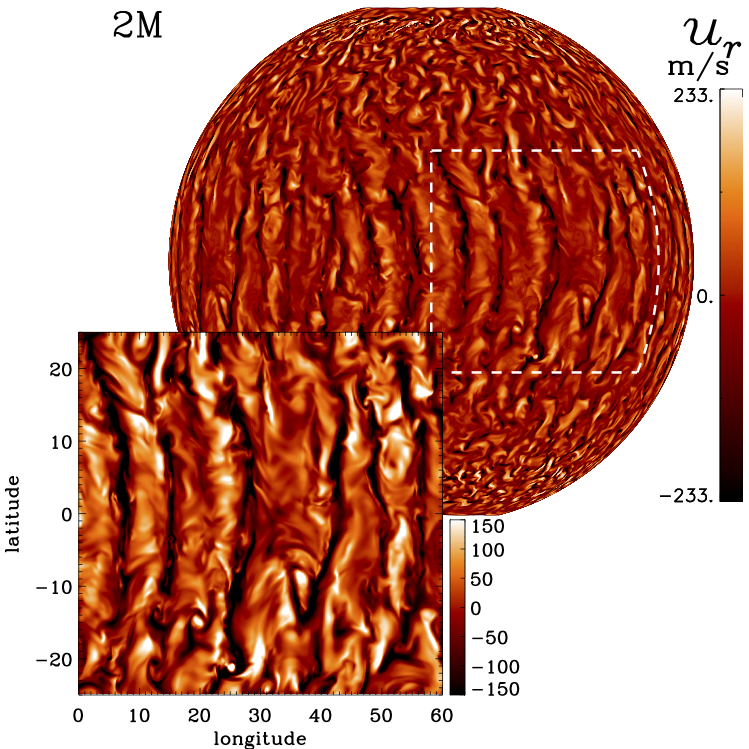}
 \includegraphics[width=0.32\textwidth]{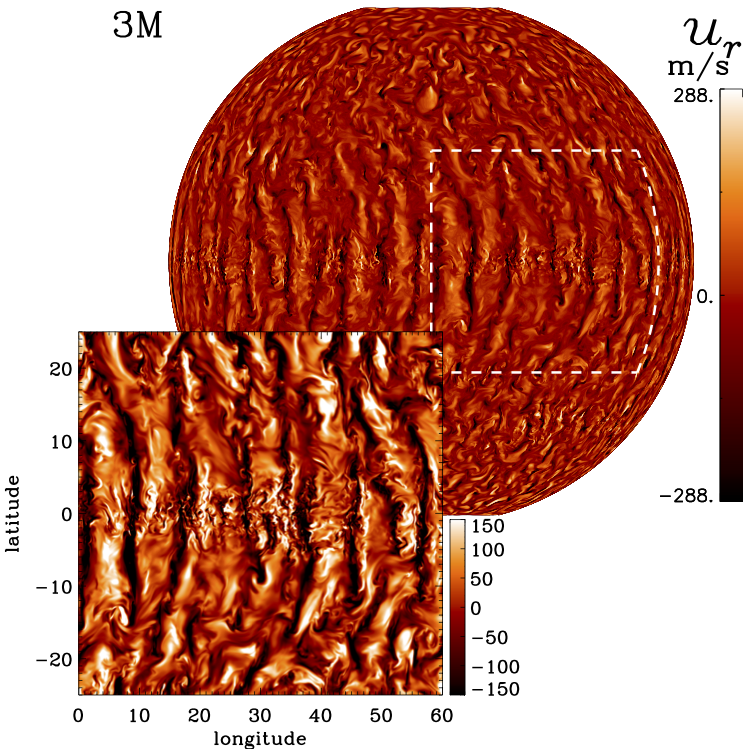}
 \includegraphics[width=0.32\textwidth]{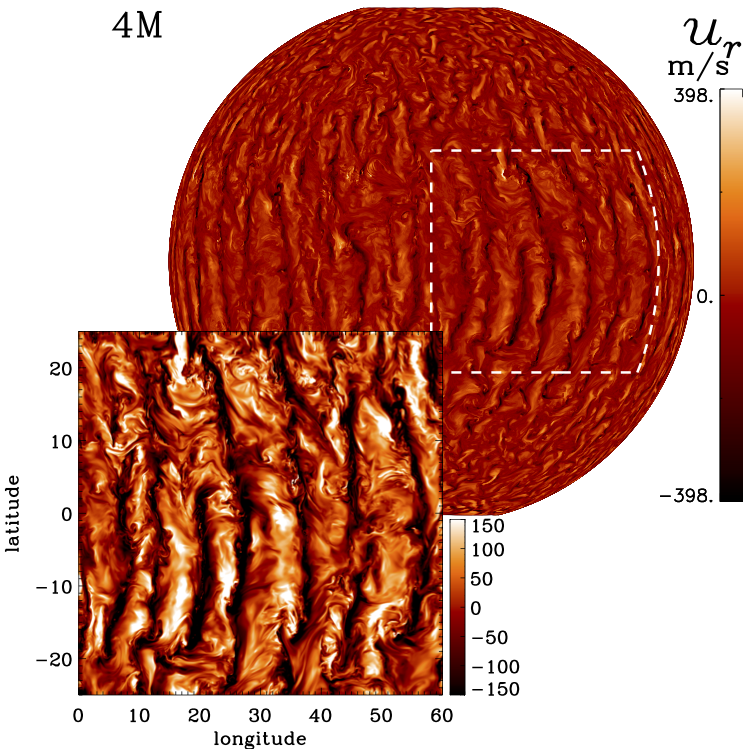}
\includegraphics[width=0.32\textwidth]{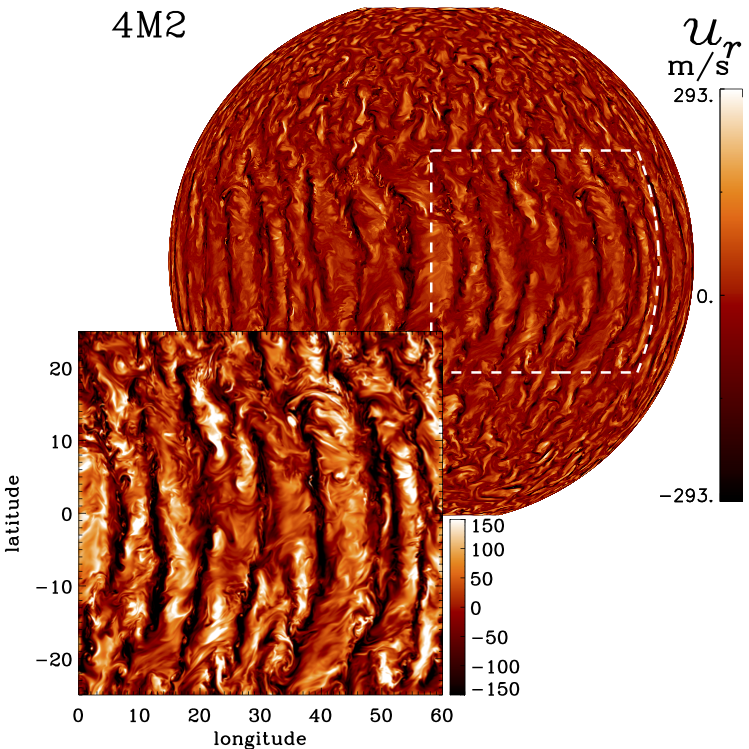}
 \end{center}
 \caption{
 Radial velocity $u_r$ at  $r=0.98\, R$  with a
 low-latitude cut-out, for all M runs in the saturated stage. The wedge is duplicated to
 form a half sphere.
 }\label{ur_slices}
\end{figure*}

\begin{figure*}[t!]
 \begin{center}
   \includegraphics[width=0.32\textwidth]{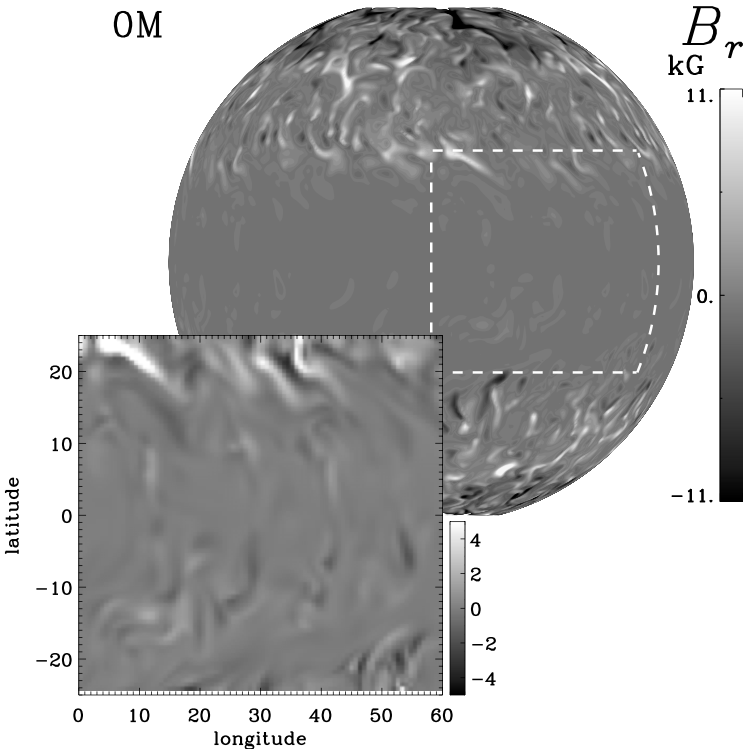}
 \includegraphics[width=0.32\textwidth]{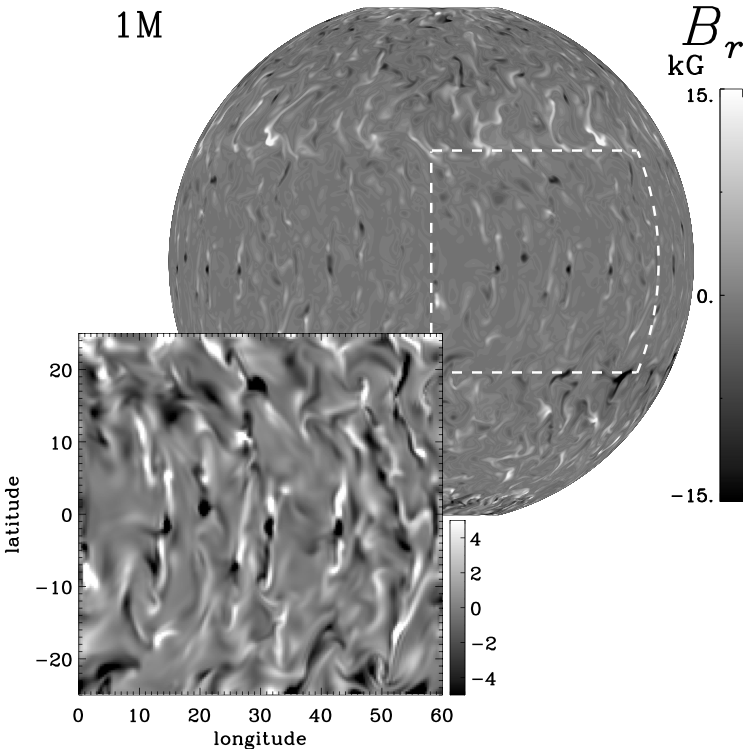}
 \includegraphics[width=0.32\textwidth]{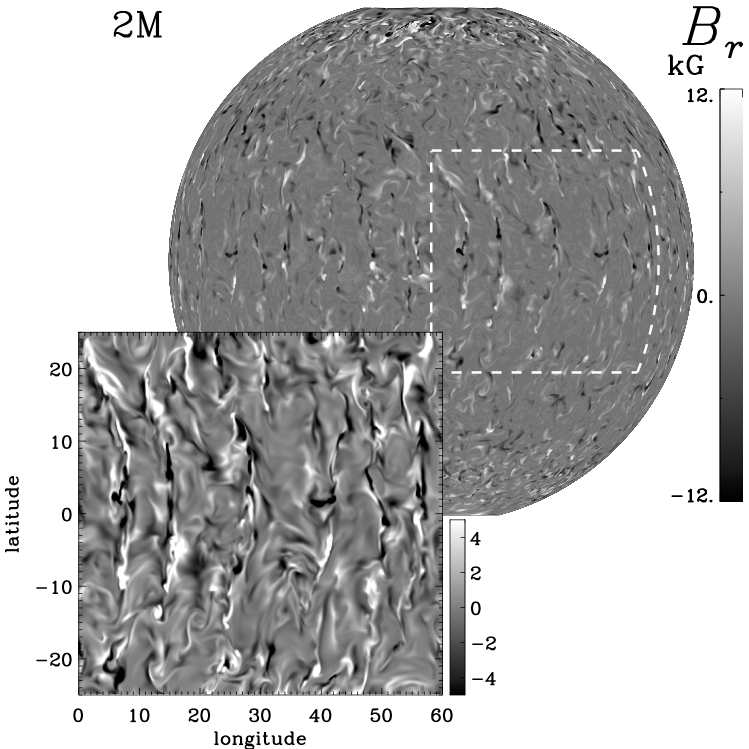}
 \includegraphics[width=0.32\textwidth]{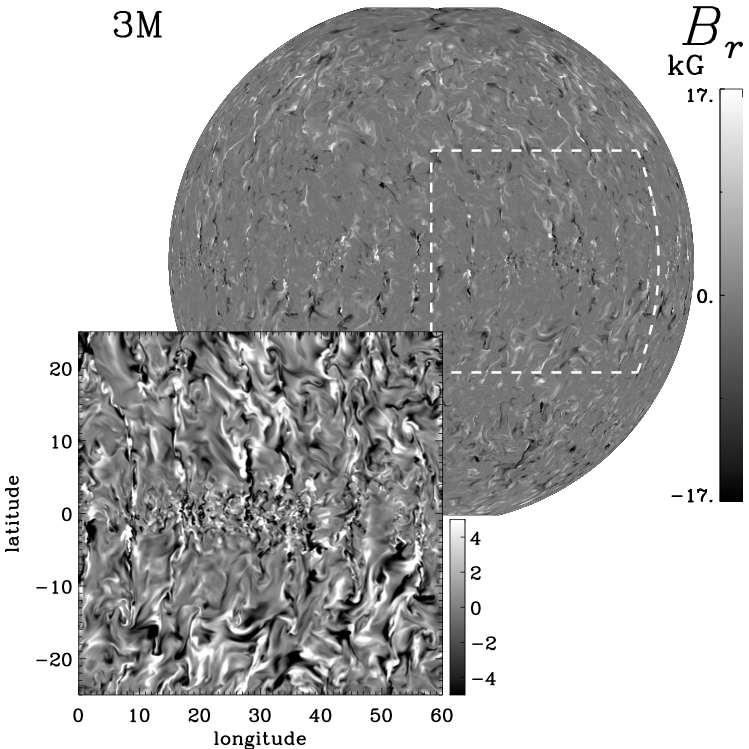}
 \includegraphics[width=0.32\textwidth]{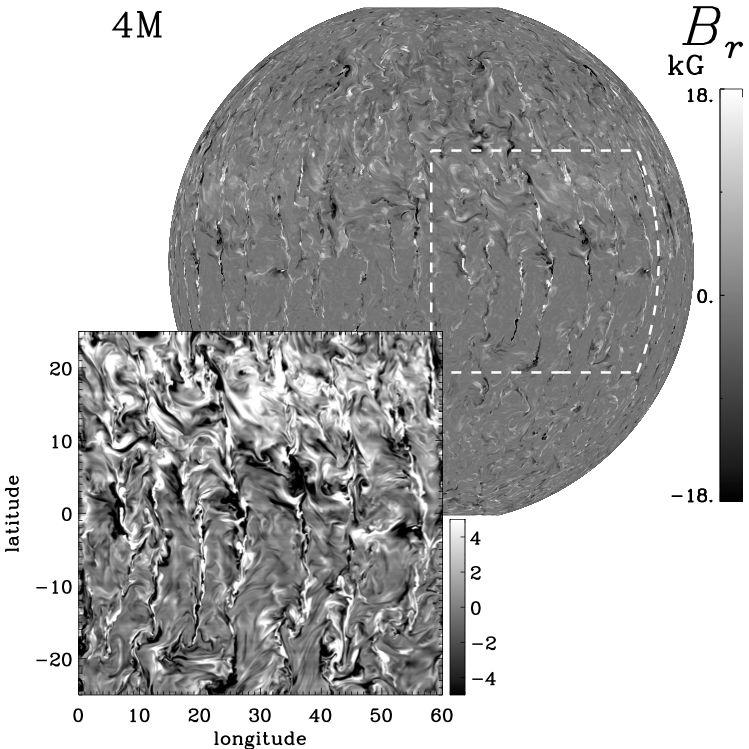}
 \includegraphics[width=0.32\textwidth]{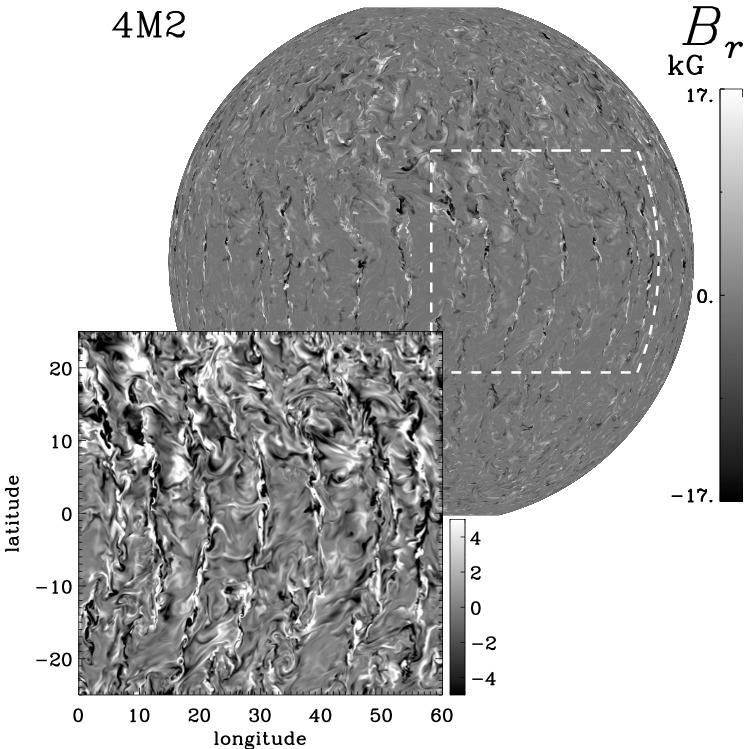}
 \end{center}
 \caption{
 Radial magnetic field $B_r$ at $r=0.98\, R$ for all M runs.
Otherwise as \Fig{ur_slices}.
 }\label{br_slices}
\end{figure*}

\begin{figure*}[t!]
 \begin{center}
 \includegraphics[width=0.32\textwidth]{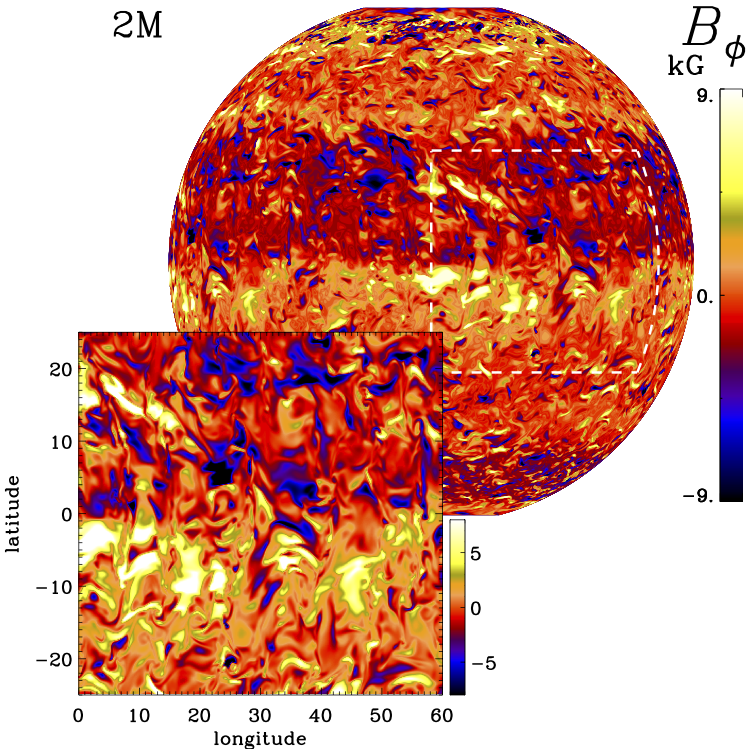}
 \includegraphics[width=0.32\textwidth]{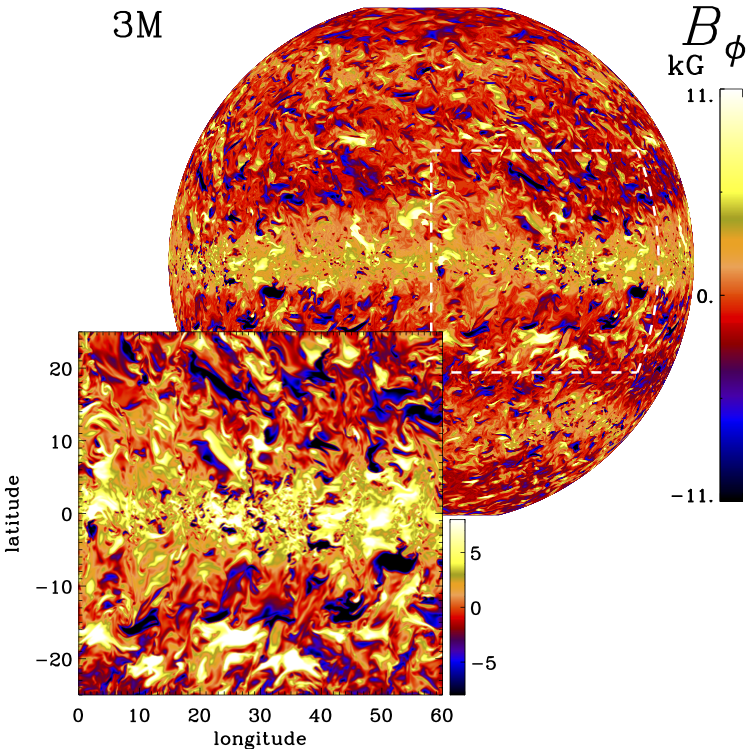}
 \includegraphics[width=0.32\textwidth]{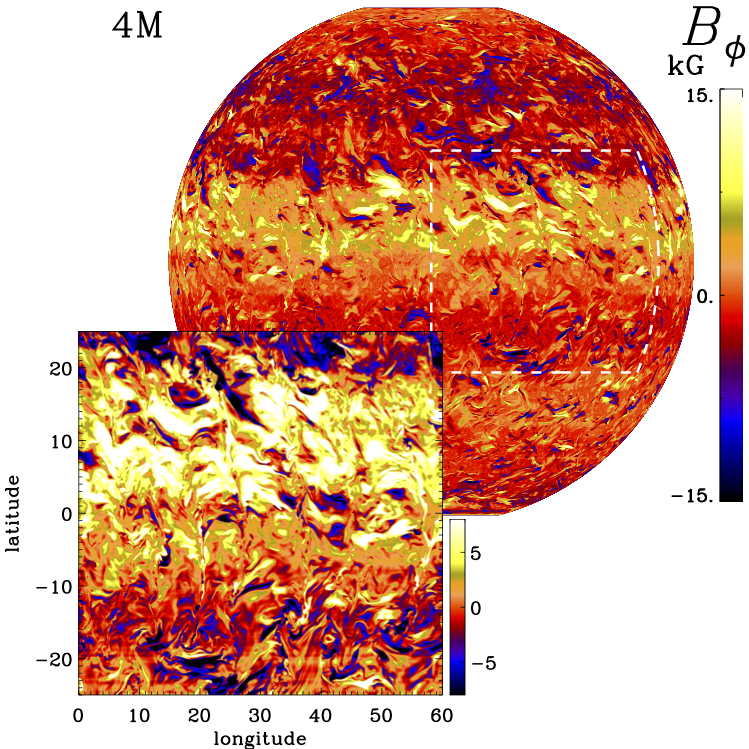}
 \includegraphics[width=0.32\textwidth]{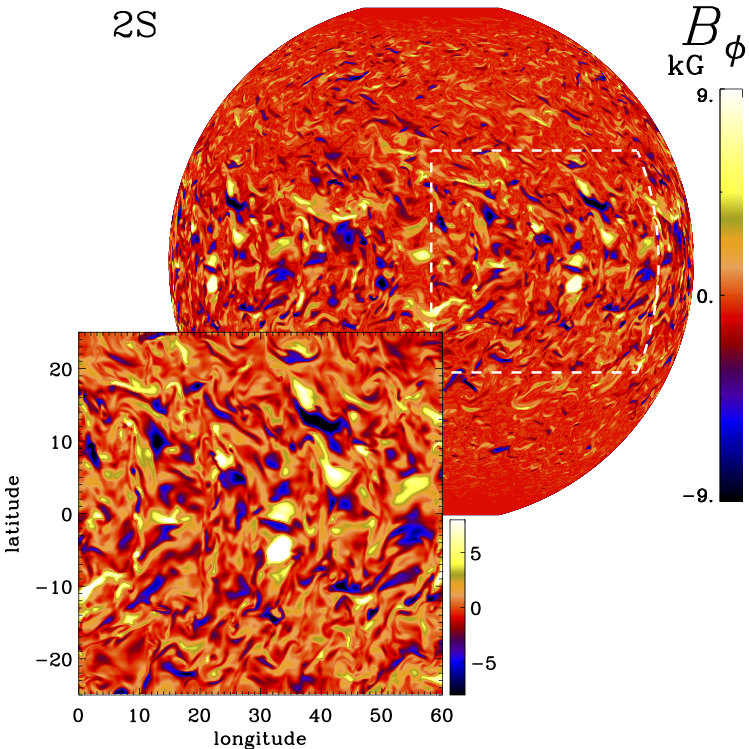}
 \includegraphics[width=0.32\textwidth]{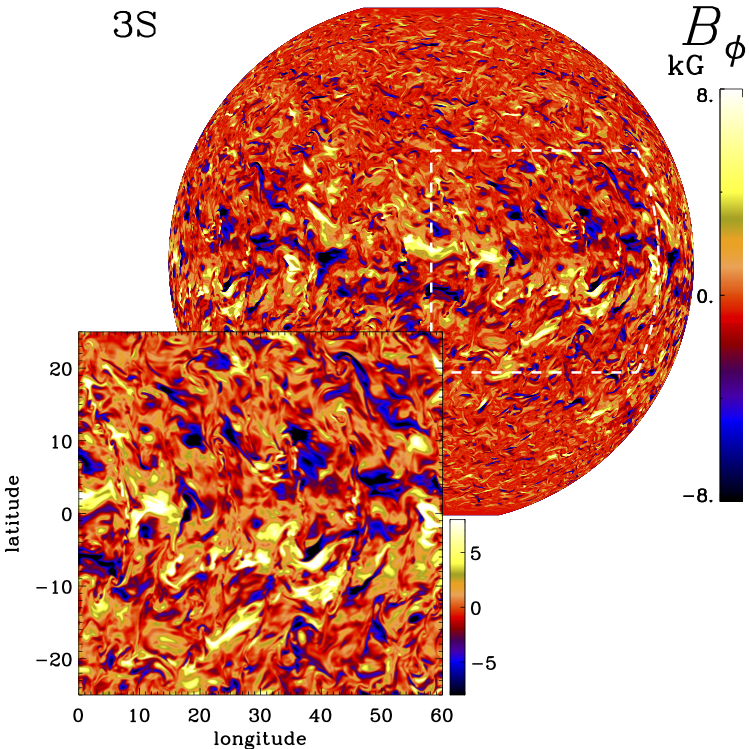}
 \includegraphics[width=0.32\textwidth]{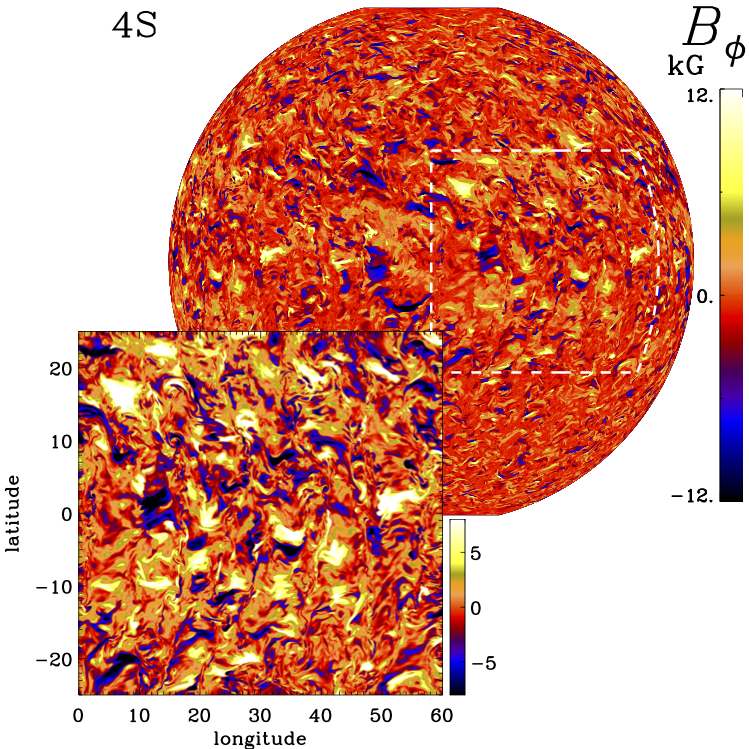}
 \end{center}
 \caption{
Azimuthal magnetic field $B_\phi$ at  $r=0.98\, R$
 for all M runs with an SSD and the corresponding S runs.
Otherwise as \Fig{ur_slices}.
 }\label{bp_slices}
\end{figure*}

The plasma is heated at the bottom by a constant heat flux
and cooled at the top by black-body radiation.
The velocity $\uu$ is stress-free on all radial and latitudinal
boundaries, and the entropy $s$ has zero derivatives at the latitudinal
boundaries. At the lower
radial and at the latitudinal boundaries, we choose
the magnetic field $\BB$ to follow a perfect conductor while being
purely radial at the top.
In the $\phi$ direction,
the boundary conditions for all quantities are periodic.

Our runs are defined by the following non-dimensional input
parameters:
We define a normalized angular frequency and the Taylor number
\begin{equation}
\Ot=\Omega_0/\Omega_\odot, \quad \Ta=\left[2\Omega_0 (0.3R)^2/\nu\right]^2,
\end{equation}
where $\Omega_\odot=2.7\times10^{-6} \s^{-1}$ is the rotation rate of
the Sun,
while the thermal, SGS-thermal, and magnetic Prandtl numbers
are\begin{equation}
\Pr={\nu\over\chi_{\rm m}}, \quad\PrSGS={\nu\over\chiSm},\quad \Pm={\nu\over\eta},
\end{equation}
where 
$\chi_{\rm m}=K(r=r_{\rm m})/\rho c_{\rm P}$ is the thermal diffusivity based on
Kramers' opacity in the middle of the convection zone ($r=r_{\rm m}$), see \Eq{eq:Kkram},
and $c_{\rm P}$ is the specific heat capacity at constant pressure.
As $K$ depends on local density and temperature, we use a
one-dimensional hydrostatic model to determine $K$ and $\rho$ for
$\Pr$.
We note that $\Pr$ in the saturated stage can be significantly
different from
the hydrostatic model.
Additionally, we define two Rayleigh numbers calculated from the same
hydrostatic  model. One is based on the Kramers heat
diffusivity $\chi$
\begin{equation}
\Ra_{\rm Kram}(r)\!=\!\frac{GM(0.3R)^4}{\nu\chi(r) R^2}
  \bigg(-\frac{1}{c_{\rm P}}\frac{{\rm d}s_{\rm hs}}{{\rm d}r}
  \bigg)\,,
\label{equ:Rakram}
\end{equation}
the other on the SGS heat diffusivity $\chiS$
\begin{equation}
\Ra_{\rm SGS}(r)\!=\!\frac{GM(0.3R)^4}{\nu\chiS(r) R^2}
  \bigg(-\frac{1}{c_{\rm P}}\frac{{\rm d}s_{\rm hs}}{{\rm d}r}
  \bigg)\,,
\label{equ:Rakram}
\end{equation}
where $s_{\rm hs}$ is the specific entropy in the hydrostatic model,
$G$ is the gravitational constant, and $M$ is the total mass of the
star.
Our specific choices of $\Ra$ reflect  the difficulty in defining
a meaningful value for fully compressible convection. To meet the
requirement of being determined
by input parameters, that is,
in the absence of convection, we use a 1D version of our setup and allow it to 
equilibrate to the hydrostatic state $s_{\rm hs}$.
As the Rayleigh numbers strongly depend on $r$ and are not always
positive in the middle of the domain (as in the non-Kramers runs), we
average them over their logarithmized positive contribution
$\ln\Ra_+$:
\begin{equation}
\mean{\Ra}=\exp\,\brac{\ln\Ra_+(r)}_r,
\end{equation}
where the subscript
$r$ marks radial averaging.

To estimate the supercriticality, we define a further
Rayleigh number, in which
the reduction of supercriticality due to rotation is compensated
by the Taylor number
$\widetilde{\Ra}=\mean{\Ra}/\Ta^{2/3}$, following \citep[e.g.][]{Ch61,R68,BTCSA23}.

We further characterize our simulations by the fluid and magnetic Reynolds numbers together with the
Coriolis number
\begin{equation}
\Rey=\frac{\urms}{\nu \kf},\quad \Rm=\frac{\urms}{\eta \kf},\quad
\Co={2\Omega_0\over\urms\kf},
\end{equation}
where $\kf=2\pi/0.3R\approx21/R$ is an estimate of the wavenumber of
the largest eddies in the domain and $\urms=\sqrt{(3/2)\brac{u_r^2+u_\theta^2}_{r\theta\phi t}}$ is
the rms velocity; the subscripts indicate averaging over $r$,
$\theta$, $\phi$ and a time interval covering the saturated state.
This definition is a good estimate of the turbulent velocity,
because the meridional circulation is weak in all of our runs and
the 3/2 factor accounts for the  omitted azimuthal component of $\uu$,
see also \cite{KMCWB13} for more details.

The non-dimensional input and solution
parameters are given
in \Tab{runs} for all runs.
In summary, we use the same model as in \cite{KKOBWKP16} except that
we apply here Kramers heat conductivity instead of a fixed
conductivity profile.

For our analysis throughout the paper, we decompose each field into a mean (axisymmetric)
and a fluctuating part, which are indicated by an overbar and  a
prime, respectively, for example, $\BB=\meanBB+\fluc{\BB}$ and
$\uu=\meanuu+\fluc{\uu}$.
Restricting to fluctuating fields, we define an $r$- and $\theta$-dependent turbulent
rms velocity
$\urmsp(r,\theta)={\left\langle\,\overline{{\bm
        u}^{\prime\,2}}\,\right\rangle_t}{}^{\!\!1/2}$, turbulent
rms magnetic field strength $\brmsp(r,\theta)={\left\langle\,\overline{{\bm
        B}^{\prime\,2}}\,\right\rangle_t}{}^{\!\!1/2}$ and turbulent
equipartition field strength
$\Beq(r,\theta)=\urmsp(\mu_0\meanrho)^{1/2}$,
where $\mu_0$ is the magnetic vacuum permeability.
The total kinetic energy density is defined as
\begin{equation}
E_{\rm kin}^{\rm tot}=\half\bbra{\rho\uu^2}_V,
\label{eq:ene1}
\end{equation}
which is further decomposed into the energy densities of the fluctuating velocity, the
differential rotation, and the meridional circulation:
\begin{equation}
E_{\rm kin}^{\rm flu} =\half\bbra{\rho\uu^{\prime\, 2}}_V\!, \: E_{\rm kin}^{\rm dif}=\half\bbra{\rho\mean{u_\phi}^2}_V\!,\: E_{\rm
  kin}^{\rm mer} =
\half\bbra{\rho\left(\mean{u_r}^2+\mean{u_\theta}^2\right)}_V.
\label{eq:ene2}
\end{equation}
 Here, $\bbra{}_V$ indicates volume averaging.
In a similar way, the total magnetic energy density
is defined as
\begin{equation}
E_{\rm mag}^{\rm tot}={\textstyle\frac{1}{2\mu_0}}\bbra{\BB^2}_V,
\label{eq:ene3}
\end{equation}
and can be split
into the energy density of the fluctuating field,
along with those of the toroidal and poloidal
mean fields:
\begin{equation}
E_{\rm mag}^{\rm flu}\!=\!{\textstyle\frac{1}{2\mu_0}}\bbra{\BB^{\prime\,2}}_V\!, \:
E_{\rm mag}^{\rm tor}\!=\!{\textstyle\frac{1}{2\mu_0}}\bbra{\meanB_\phi^2}_{\!V}\!,\:
E_{\rm mag}^{\rm pol}\!=\!{\textstyle\frac{1}{2\mu_0}}\bbra{\meanB_r^2+\meanB_\theta^2}_V\!\!.\hspace{-5mm}
\label{eq:ene4}
\end{equation}

The presented analyses and quantities
are performed and
calculated from the saturated stage of the simulations.
Rescaling to physical units
is based on the solar rotation rate $\Omega_\odot=2.7\times10^{-6} \s^{-1}$,  solar
radius $R=7\times10^{8} \m$, density at the bottom
of the domain $\rho(0.7R)=200 \kg/\m^3$, and
$\mu_0=4\pi\cdot10^{-7}$~H~m$^{-1}$.
As discussed in \cite{KMCWB13} and \cite{KMB14}, this is
one particular choice,  giving meaningful results for the input flux, the
velocities, and magnetic field strengths in the bulk of the convection
zone.
All simulations were performed using the {\sc Pencil Code} \citep{PC21}.

\section{Results}
\label{sec:results}

Our goal is to study the LSD and SSD together
in a setup including
a physically motivated heat conductivity based on Kramers'
opacity.
As our starting point, we use the model of \cite{KKOBWKP16}, with the only
difference that the prescribed
heat conductivity is replaced by a Kramers-opacity based one;
this represents Run 0M. We then lower, step by step, the
viscosity $\nu$, the magnetic diffusivity $\eta$ and the SGS heat
diffusivity $\chiS$ making the simulation gradually more turbulent while
keeping $\PrSGS$ and $\Pm$ constant.
At each step, indicated
by the number in the run label, the diffusivities are
halved, leading to a total reduction by factor of 16 from
Set 0 to Set 4.
Technically, this requires doubling the resolution and remeshing
the run at each step, and eventually
running the simulation in the saturation regime for a sufficient time span.

To study the effects of the SSD in isolation and to check whether it
is indeed present, we fork each MHD
run in two, having identical setups:
In the  first one, denoted with `S", we
remove the mean field $\meanBB$ at
every fifth\footnote{This cadence was chosen to avoid slowing down the
  computing while still removing the large-scale field efficiently.}
time step,
hence no LSD can develop.
When an SSD is sustained, we study it in detail.
In the second one, denoted by ``M", 
mean-field removal  
is not applied, hence the LSD can develop freely.
Finally, we also perform
the corresponding
hydrodynamic simulations, denoted by ``H".
Note that Run 4M2 is basically the same  as 4M except that we have not
re-meshed and re-started from 3M, which has an LSD, but
re-started from 4S, where only an
SSD is present, yet
allowing for the growth of the LSD after the re-start.
Our motivation was here to study whether an LSD can be excited
and grow in the presence of already existing strong
magnetic fluctuations.
Runs 0M to 2M are similar to Runs G1 to G3 in \cite{KKOWB17},
where the difference is in the use of the Kramers-based heat conductivity
in our work and in a normal magnetic field condition at the latitudinal boundaries in G2 \& G3.

All runs are listed in  \Tab{runs} with their 
control and solution parameters.
We note that an SSD is present in Sets 2--4, implying that its
critical magnetic Reynolds number
lies between roughly 60 and 130.
This is consistent with the study of \cite{KKOWB17}, where an SSD is
typically found for $\Rm>60$.
Interestingly, run G2 of \cite{KKOWB17} has an SSD with $\Rm=66$
in contrast to our Run 2M with $\Rm=61$ that does not excite an SSD.
Either the SSD is very close to critical or the slight differences in the
setups are causal.
 We note here that $\Rm$ in both their
and our work 
is only an average and hence does not
reflect the detailed local dynamics.
The critical $\Rm$ found in our work is somewhat higher than
what is obtained from theoretical models for smooth velocity fields
with low compressibility, yielding predictions of the order of
30--60 \citep{BS05}, and 
$\Rm=30-50$ from simple isothermal forced
turbulence models \citep[e.g.][]{SHBCMM05,VPKRSK21,WKGR23}.

Decreasing the diffusivities leads to a strong increase in the
Rayleigh numbers $\mean{\Ra}_{\rm Kram}$ and $\mean{\Ra}_{\rm SGS}$
by a factor of more than 100. However, as also the Taylor number, $\Ta$,
increases significantly by a factor of nearly 300, we need to look at
the compensated Rayleigh number $\widetilde{\Ra}_{\rm SGS}$ to assess
whether or not the supercriticality increases.
Indeed, $\widetilde{\Ra}_{\rm SGS}$ is nearly 4 times larger in the
run with the 
lowest diffusivities than in the one with the highest.
The rotational influence on the convection in terms of $\Co$
decreases only slightly when the 
diffusivities are decreased
because the
turbulent convection becomes slightly stronger.
$\Pr$ is above unity for Sets 0 and 1 and below unity for Sets 2--4,
indicating that heat conduction (in the middle of the convection
zone) is dominated by the SGS contribution for Sets  0 and 1 and by the
Kramers contribution otherwise,
however, one needs to keep in mind that these two terms 
involve different gradients.

All runs reach dynamical saturation in terms of their running time
$t_{\rm max}$ being a multiple of the convective turnover time
$\tau=1/\urms\kf$; i.e. $t_{\rm max}/\tau > 3000$ for Sets 0, 1 \& 2,
$t_{\rm max}/\tau > 360$ for Set 3 and $t_{\rm max}/\tau > 2.4$ for
Set 4. In terms of the turbulent magnetic diffusion time $\tau_{\rm
  mag}^{\rm turb}=(0.3R)^2/\eta_{t}$ with $\eta_t=\urms\kf/3$, which
is important for the evolution of the mean field, all runs from Sets
0–3 reach multiple $\tau_{\rm mag}^{\rm turb}$ and hence can be
considered to be in a steady state; see also the discussion
in \Sec{sec:eng}.

\subsection{Overview of the dynamics}

As shown in \Fig{ur_slices} for all the M runs,
the radial velocity becomes more turbulent
and develops progressively more small-scale structures
when lowering  the diffusivities, hence increasing the Reynolds
numbers. For all runs, prominent thermal Rossby waves, also known as
Busse columns or banana cells
\citep[e.g.][]{Busse70,Busse76,FH16,BCG22} are present outside the
tangent cylinder in the equatorial regions,
in agreement with earlier
models and theoretical expectations.
Interestingly, the longitudinal degree of these
waves does not vary much with increasing Reynolds number, being mostly
between $m=32$ and $m=40$ for all runs;
see \App{sec:ban} for a more detailed analysis.

In \Fig{br_slices}, we show the corresponding radial magnetic
field close to the surface for all the M runs.
Near the equator, the radial magnetic field is mostly concentrated in
the downflow lanes of the banana cells
where it forms complex structures
with small bipolar patches.
The radial field is mostly dominated by
magnetic fluctuations, while the mean field is not visible near the equator.
At higher latitudes in Runs 0M--2M, one can see hints of longitudinal bands
of same polarity tracing the weak mean radial field. For 3M, 4M and 4M2, these
patterns are no so clearly visible anymore.

In \Fig{bp_slices}, we show the azimuthal magnetic field
for Runs 2M, 2S, 3M, 3S, 4M, and 4S.
For this component, the difference between
Sets S and M is most pronounced. In the M runs, the mean field is
clearly visible at all covered
latitudes.
The magnetic fluctuations in these runs are also present ubiquitously.
In Run 2S, the fluctuating magnetic field is
more concentrated near the equator.
The fluctuations become distributed over
progressively wider latitude ranges when the
Reynolds numbers are increased, see Runs 3M to 4M.
Interestingly, the banana cell pattern
does not produce a strong imprint on
the azimuthal magnetic field structure.

\begin{table*}[t!]\caption{
Energy densities for all runs.
}\vspace{0pt}\centerline{\begin{tabular}{l|rrcccccccc}
Run &$E^{\rm tot}_{\rm kin}$\phantom{5}&$E^{\rm dif}_{\rm kin}$\phantom{5}&$E^{\rm mer}_{\rm kin}$&$E^{\rm flu}_{\rm kin}$&$E^{\rm tot}_{\rm mag}$&$E^{\rm tor}_{\rm mag}$&$E^{\rm pol}_{\rm mag}$&$E^{\rm flu}_{\rm mag}$&$E^{\rm tot}_{\rm mag}/E^{\rm tot}_{\rm kin}$&$E^{\rm flu}_{\rm mag}/E^{\rm flu}_{\rm kin}$\\[.8mm]
\hline
\hline\\[-2.5mm]
0M       & 2.225&1.562&0.006&0.658&0.369&0.072& 0.048&0.250&0.166&0.379\\
0H     &17.089&16.159&0.010&0.920\\
\hline
1M          &1.267&0.546&0.006&0.716&0.604&0.172&0.047&0.385&0.476&0.538 \\
1H      &  4.397&3.457&0.010&0.930 \\
\hline
2M          &0.933&0.213&0.005&0.715&0.763&0.198&0.044&0.521&0.819&0.729\\
2S           &5.643&4.503&0.010&1.130&0.140&0.000&0.000&0.140&0.025&0.124\\
2H         & 8.697&7.514&0.012&1.171\\
\hline
3M          & 0.923&0.197&0.005&0.721&0.510&0.057&0.023&0.430&0.553&0.597\\
3S          & 1.448&0.542&0.007&0.900&0.228&0.000&0.000&0.228&0.158&0.254\\
3H         & 7.352 &6.347&0.011&0.993\\
\hline
4M          & 0.854&0.144&0.005&0.705&0.876&0.190&0.030&0.656&1.026&0.931 \\
4M2        & 1.016&0.132&0.006&0.879&0.604&0.009&0.006&0.590&0.594&0.671 \\
4S           & 1.004&0.137&0.005&0.862&0.514&0.000&0.000&0.515&0.512&0.597 \\
\hline
\hline
\end{tabular}}\label{runs2}\tablefoot{
The energy densities
(columns 2 to 9)
are  in $10^5$ J/m$^3$
and their definitions are given in \Eqss{eq:ene1}{eq:ene4}.
}
\end{table*}

\begin{figure}[t!]
\begin{center}
\includegraphics[width=0.965\columnwidth]{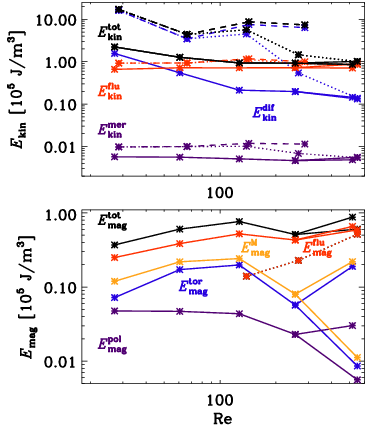}
\end{center}\caption[]{
Dependence of kinetic and magnetic energies on 
Reynolds number $\Rey$. In the top panel we show the total kinetic energy $E_{\rm kin}^{\rm tot}$ (black), which is
composed of the energy of the fluctuating flow $E_{\rm kin}^{\rm
  flu}$ (red), the differential rotation $E_{\rm kin}^{\rm dif}$ (blue), and
the meridional circulation $E_{\rm kin}^{\rm mer}$ (purple) for
H runs (dashed), M runs (solid) and S runs (dotted).
Bottom: total magnetic energy density $E_{\rm mag}^{\rm
  tot}$ (black) composed of the energy densities of the fluctuating
magnetic field $E_{\rm mag}^{\rm flu}$ (red), the mean toroidal
$E_{\rm mag}^{\rm tor}$ (blue), and mean poloidal magnetic field $E_{\rm
  mag}^{\rm pol}$ (purple).
$E_{\rm mag}^{\rm M}=E_{\rm mag}^{\rm tor} + E_{\rm mag}^{\rm pol}$
is the energy density of the mean magnetic field
(yellow);
see \Eqss{eq:ene1}{eq:ene4} for definitions.
The lower mean-field energy densities
for $\Rey\sim530$ indicate Run~4M2, which has been
started from Run~S4 and is possibly not yet saturated, hence the bifurcation.
}\label{energies}
\end{figure}

\subsection{Energies}
\label{sec:eng}

Next we look at the total (volume-averaged)
energy densities as shown in
\Fig{energies} and \Tab{runs2}.
For the H runs, the kinetic energy is generally higher
than for the other sets, and it is dominated by the differential
rotation at all Reynolds numbers.
For the M runs, the contribution of the differential
rotation becomes weaker and sub-dominant
w.r.t the velocity fluctuations with increasing Reynolds numbers.
For the 
lowest diffusivities
investigated,
however, the contribution of differential rotation
to the total kinetic energy seems to level off at a value of
roughly 15\%.
The contribution of the meridional circulation is tiny for all runs.
The energies of the fluctuating velocities
remain for all sets roughly constant at increasing Reynolds
numbers  being, however, a bit higher for the H runs than for
the M runs.

The S runs with small $\Rey$  show similar dominance of differential rotation, but its
contribution diminishes for growing $\Rey$ similarly
to the M runs. The S runs show basically a transition from an
almost
purely hydrodynamic
state (2S) to a fully magnetically dominated state (4S).
This has to be attributed to the SSD, 
strengthening
strongly with growing Reynolds numbers, as will be discussed next.

The magnetic energy increases with increasing $\Rey$ for the first
three M runs. This is mostly due to the increase of the mean field.
For higher $\Rey$, the energy in the mean field actually decreases,
whereas the contribution of the small-scale field increases.
The results for the 
lowest diffusivities
(4M, 4M2) need to be taken with caution:
Run 4M has been restarted from a remeshed earlier stage of 3M, at which the
mean-field energy was still high, see \Fig{but} and its discussion in
\Sec{sec:magfield}.
Run~4M was most likely not run long enough to let
the magnetic field reach a new saturated stage. For 3M, it took
already $\sim \!10~{\rm yr}$ to saturate.
Run 4M2 has been started from 4S to see how fast the mean field can
recover after it has been removed. After running for
$\sim\! 0.06~{\rm yr}$, the
mean field was still very weak.

For all runs, the mean field is  dominated by its toroidal part.
For all M runs, the small-scale field contribution dominates the total
magnetic energy and shows some tendency to increase for high $\Rey$.
In the runs without SSD (0M and 1M), the fluctuations 
contribute by roughly 65\% to the total magnetic energy.
When the SSD starts operating, this contribution increases from
 68\% (Run 2M) to 75\% (4M).

In the S runs, we observe a significant increase of the magnitude
of the magnetic fluctuations due to the
strengthening SSD.
Let us assume that $E_{\rm mag}^{\rm flu}$ in the S runs gives a
good representation of the strength of the SSD-generated field 
also in the
corresponding M runs, despite some differences in the flow dynamics.
Then for $\Rey\sim130$ a quarter of $E_{\rm mag}^{\rm flu}$ is due to the SSD,
increasing to nearly 80\% for $\Rey\sim500$.

In our interpretation, at low $\Rey$, the quenching of the differential rotation
is caused by the small-scale field generated by the tangling of the
large-scale magnetic field, rather than by the large-scale field directly.
At higher $\Rey$, the SSD also generates small-scale field, leading
to a further quenching of the differential rotation.

If we do not consider Run~4M, given the issues mentioned above,
the total magnetic field energy tends to saturate at high $\Rm$
because the increase in the fluctuating-field energy
compensates for the decrease in the mean-field magnetic energy.
This is in contrast to previous studies
\citep{NBBMT13, HRY16, KKOWB17}, where the total field shows a steady
increase with $\Rm$.
However, as far as the mean field is concerned,
our study is roughly consistent with the results of \cite{NBBMT13} and
\cite{HRY16} 
as these authors find that
the mean-field energy decreases with $\Rm$; yet, our study is
inconsistent with our previous work \citep{KKOWB17},
where the mean field does not decrease.

The total magnetic energy is close to equipartition 
with the total kinetic energy only at the highest $\Rey$ (4M) due to the strong
mean-field and its tangled fluctuating field, see last two columns of
\Tab{runs2}.
In the pure SSD runs, the field reaches only 50\% of the equipartition value. This does not
agree with the results of HKS22, where
super-equipartition fields were found at
their highest resolution.
Comparison with recent work by \cite{YanCalc22} on large- and
small-scale convective dynamos in plane layers, yields several
differences:  While they find the ratio of magnetic to kinetic energy
falling with increasing $\widetilde{\Ra}$, yet showing some tendency 
to saturation from  $\widetilde{\Ra}\approx 50$ on, we see the
opposite trend, yet in the $\widetilde{\Ra}$ interval
$[48,184]$. Also, we cannot confirm their conclusion that SSD are
likely to yield smaller energy ratios. One has to obey, though, that
their simulations were performed at $\widetilde{\Rm}=\Rm \Ta^{1/6}
\approx {\cal O}(1)$ whereas we covered the $\widetilde{\Rm}$ interval
$[26,210]$.

\subsection{Overshoot and Deardorff layers}
\label{sec:kramers}

Similar as in the work of \cite{KVKBS19} and \cite{VK21} we found that
using a Kramers-based heat conductivity causes the development of sub-adiabatic,
yet convective layers
in addition to the usual convective zone.
From top to bottom of the domain, the zones are defined as
\begin{alignat}{2}
\meanF_{\rm enth}>0 &\text{,}\quad \dd s/\dd r<0 \quad &&\text{buoyancy zone (BZ)}\hspace{-2mm}\\
\meanF_{\rm enth}>0 &\text{,}\quad \dd s/\dd r>0 \quad &&\text{Deardorff zone (DZ)}\hspace{-2mm}\\
\meanF_{\rm enth}<0 &\text{,}\quad \dd s/\dd r>0 \quad&&\text{overshoot zone (OZ)}\hspace{-2mm}\\
\meanF_{\rm enth}<0 &\text{,}\quad \left|\meanF_{\rm enth}\right| < 0.03
                                  \meanF_{\rm tot}\quad &&\text{radiative zone (RZ)},\hspace{-2mm}
\end{alignat}
where $\meanF_{\rm enth}=c_{\rm
  P}\overline{\fluc{(\rho u_r)}\fluc{T}}$ is the radial enthalpy
flux, and $\meanF_{\rm tot}=\meanF^{\,\rm rad} (0.7\,R)$ is the
total flux, defined by the flux through the bottom boundary.
In the definitions above,
all fluxes are additionally averaged over latitude, hence
they only depend on radius $r$.

\begin{figure}[t!]
\begin{center}
\raggedright
\includegraphics[width=0.119\textwidth]{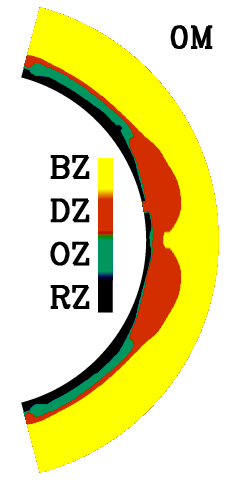}
\includegraphics[width=0.119\textwidth]{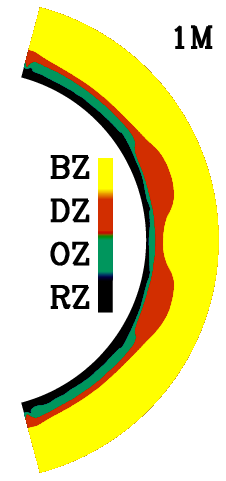}
\includegraphics[width=0.119\textwidth]{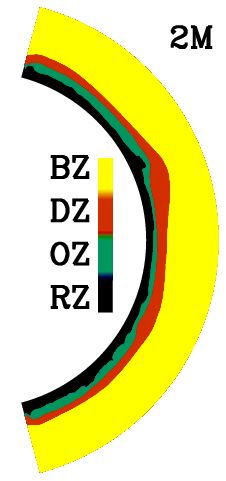}
\includegraphics[width=0.119\textwidth]{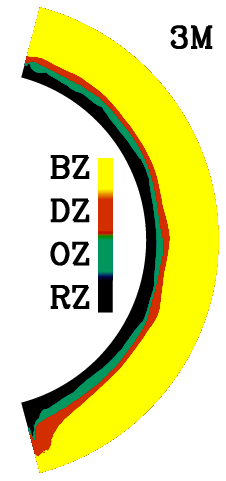}\\
\hfill
\includegraphics[width=0.119\textwidth]{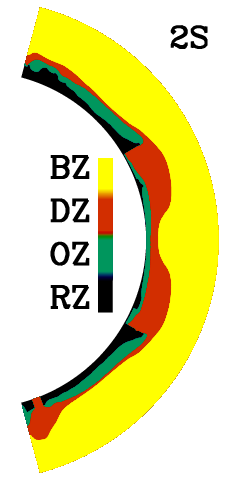}
\includegraphics[width=0.119\textwidth]{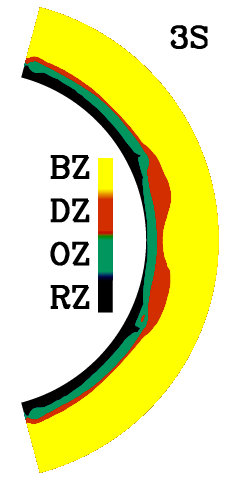}\\
\includegraphics[width=0.119\textwidth]{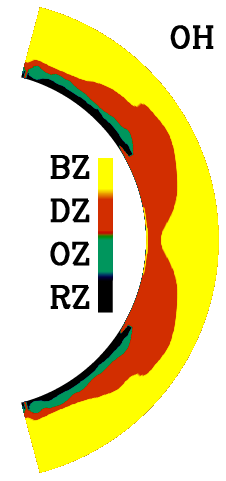}
\includegraphics[width=0.119\textwidth]{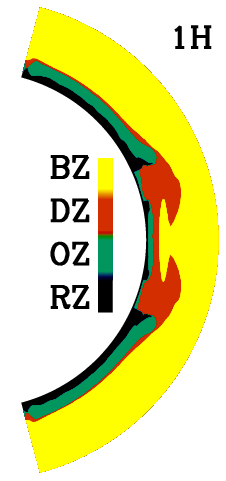}
\includegraphics[width=0.119\textwidth]{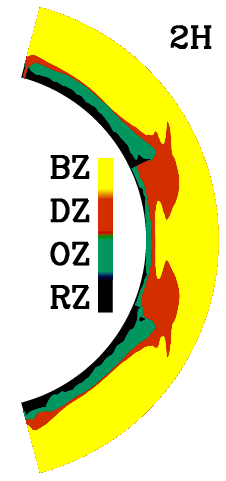}
\includegraphics[width=0.119\textwidth]{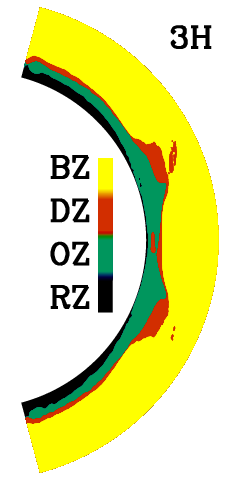}\\
\end{center}\caption{
Visualization of the four zones formed in the
simulations: buoyancy  (BZ, yellow), Deardorff  (DZ, red), overshoot
 (OZ, green), and radiative zone (RZ, dark blue), see \Sec{sec:kramers} for definitions.
}\label{convlay}
\end{figure}

We investigate the dependence of these zones on $\Rey$ for different
runs in Fig. \ref{convlay}.
As a first result, we observe that approximately
the lower quarter of the domain is convectively stable (even more so for low
$\Rey$), consistent with previous works in similar parameter regimes
\citep{KVKBS19, VK21}.

The Deardorff zone is more pronounced near the equator in the
high-diffusivity runs
runs, especially in hydro cases. As $\Rey$ increases,
the Deardorff zones become narrower near the equator and more
uniform over latitude, while their radial extent at high
latitudes depends on the presence of magnetic fields. In the M Runs,
the DZ becomes very thin and is partly replaced by a thin overshoot
layer and an extended radiative zone for high $\Rey$. In the S Runs,
the DZ is more pronounced at low latitudes, even at high $\Rey$. For
the H Runs, it is always pronounced at low latitudes; however, at high
$\Rey$, most of it is replaced by the overshoot zone.

Regarding the discussion on how the depth of the OZ depends on
$\Rey$ \citep[see][]{H2017,K2019}
our conclusion is complex. For all runs, at high latitudes the depth seems to remain
rather constant, in agreement with \cite{K2019, K2021}. However, at
low latitudes, the depth decreases with $\Rey$ for the H runs,
aligning with \cite{H2017}, but increases in the S runs.
As for the depth of the DZ, \cite{K2021} find a weak dependence
on $\Rey$, which we can confirm
for high latitudes but not throughout.

\begin{figure}[t!]
\begin{center}
\includegraphics[width=\columnwidth]{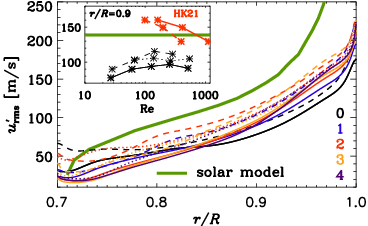}
\end{center}\caption{
Radial profile of latitudinally averaged turbulent velocity $\urmsp$ of all runs.
Colors indicate the different run sets with different $\Rey$ as
indicated.
Solid lines, M runs, dotted S runs, dashed H runs.
The thick green line indicates values from the solar.
  model of \cite{Stix:02} based on mixing-length-theory.
The inset shows $\urmsp$ at a fixed radius $r=0.9\,R$
as function of $\Rey$.
There we overplot $\urmsp$ of \cite{HK21} using $\Rey$ estimated
in HKS22 (solid red) and a re-estimate according to \Eq{ReHK22} (dashed).
}\label{urms}
\end{figure}

\begin{figure}[t!]
\begin{center}
\includegraphics[width=\columnwidth]{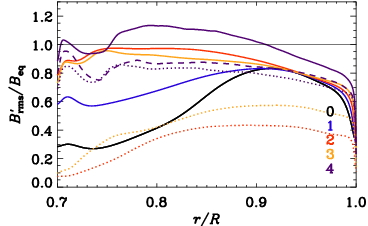}
\end{center}\caption{
Radial profile of the normalized latitudinally averaged turbulent magnetic field
$\brmsp/\Beq$ of all runs. Color coding and line style as in
\Fig{urms}, with the exception of Run
4M2 (dashed purple).
}\label{brms}
\end{figure}

\subsection{Flow and magnetic field distribution}

The variation
of the fluid and magnetic Reynolds numbers also impacts
the distribution of turbulent velocity and magnetic fields. In
Fig. \ref{urms}, we illustrate the latitudinally averaged radial
profiles of $\urmsp$ for all runs
and compare  with the values for the Sun, relying on
mixing-length theory (MLT),
see the green line for
the Standard Solar Model of Table
6.1 in \cite{Stix:02}.
As MLT  provides only radial velocities, we have multiplied them with
$\sqrt{3}$ to get an estimate of $\urms$ assuming isotropy of the
flow.
These values of the convective velocities 
need to be taken with caution as they rely on crucial assumptions,
e.g. the value of the MLT parameter. Also, helioseismic
analysis has put doubt on them
\citep[e.g.][]{HDS12}.

Firstly, our values are somewhat lower than the ones from the Sun,
which is expected due to the stronger rotational effects
 in our simulations, as indicated by the Coriolis numbers ranging
from $\Co=8$ to $9$. Near the surface, significant discrepancies
arise, likely due to the absence of
a strong density stratification. The M
runs show the poorest agreement, while the H runs exhibit higher
velocities compared to their 
M counterparts. The S runs fall
between these two. We interpret this as the turbulent velocities being
suppressed by the presence of magnetic fields, primarily by the
ones generated by the large-scale dynamo.

Upon investigating the turbulent velocity as a function of $\Rey$ at $r=0.9\,R$
(see inset in Fig. \ref{urms}), we observe
as before that for all H runs, $\urmsp$ is
larger than for the M runs, with the S runs consistently in
between. Velocity fluctuations increase for all run sets initially
with $\Rey$ and then decrease for higher $\Rey$; however, this decrease
is not as pronounced as the initial increase. The decrease in $\urmsp$
in Run 4M might be caused by the magnetic field, but we 
observe a mild decrease in Run 3H, too, and even an increase for Run 4S.
Therefore, this interpretation may not be entirely valid.

We also compare our velocities to the results of HK21, who find that
the SSD suppresses  the turbulent
velocity as their effective $\Rey$ increases. Since they do not
explicitly define $\Rey$ due to the usage of a slope-limited diffusion
(SLD) scheme without explicit diffusivities, they estimate $\Rey$
using the Taylor microscale. As an alternative, we estimate their
$\Rey$ using the
grid spacing of our Set 4:
\begin{equation}
\Rey_{\rm HK21}=\Rey_{\rm 4}\Delta\phi_{\rm 4}/ \Delta\phi_{\rm HK21}.  \label{ReHK22}
\end{equation}
Here, $\Rey_{\rm 4}=530$ is a rough average of all $\Rey$ in Set 4,
$\Delta\phi_{\rm 4}=\pi/2/2048$ the grid size in this set, and
$\Delta\phi_{\rm HK22}=2\pi/(1536, 3072, 6144)$ the variable
grid size of the three runs of HK21.
This approach
ensures that $\Rey$ does not increase much more than by a factor
of two when the resolution is doubled, in contrast to the approach of HKS22.

Comparison reveals two significant findings: Firstly, as evident
from the inset of Fig. \ref{urms}, their $\urmsp$ values are markedly
higher than ours and exceed those of the Standard Solar Model
\citep{Stix:02} for their low and medium resolutions.
Secondly, our simulations do not exhibit a
decrease in $\urmsp$ with increasing $\Rey$,
as notable as observed in HK21;
instead, we observe only a mild decrease at high $\Rey$. Two 
explanations for this behavior can be considered: Firstly, it may be
linked to the more efficient SSD in HK21's highest resolution
simulation, likely a result of higher turbulent velocities. Secondly,
it could be a consequence of their use of a pure SLD scheme for
viscous and magnetic diffusion, as opposed to the constant explicit
diffusivities employed in our study.
This is supported by the fact that HK21 achieves a more efficient SSD at an even lower resolution
than we do.

However, if the turbulent velocities were quenched by the SSD at high
Re, we would expect to see a suppression of $\urmsp$ as a function of
$\Rey$ only for M and S runs, yet we also observe it for H runs.
Another possible cause may be the specific setup of HK21, where the
energy fluxes are fixed to the solar luminosity at both radial
boundaries, forcing the total energy to remain constant at its initial
value.
Such a constraint assumes that the initial stage is already very close
to the final solution, because large departures from the initial
stratification are not possible. This makes the solution intrinsically
not self-consistent.
Presently, we are confident that our model,
allowing thermal energy to adjust to the dynamics, is more
self-consistent and hence more realistic.

\begin{figure}[t!]
\begin{center}
\includegraphics[width=0.42\textwidth]{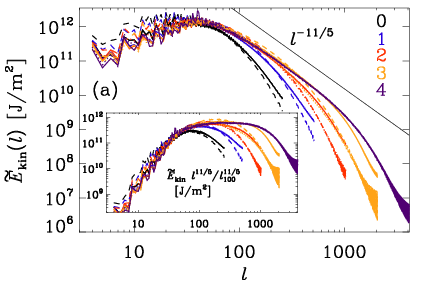}
\includegraphics[width=0.42\textwidth]{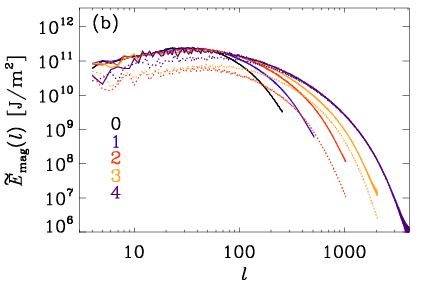}
\includegraphics[width=0.42\textwidth]{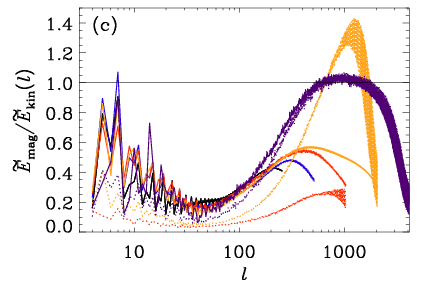}
\includegraphics[width=0.42\textwidth]{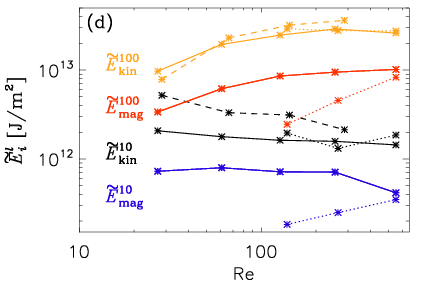}
\end{center}\caption{
Spectra of kinetic (a) and magnetic (b) energy
along with their ratio (c) near the surface, $r=0.98\,R$,
all excluding the $m=0$
contribution.
As before: M runs - solid, S runs - dotted, H runs - dashed lines,
while different
colors indicate run sets with distinct $\Rey$.
Panel (d) highlights the $\Rey$ dependence of kinetic and magnetic
spectra for $l=10$ and $100$.
}\label{spectra}
\end{figure}

Next, we examine the latitudinally averaged turbulent magnetic field
strength $\brmsp$, normalized by the equipartition field strength
(see \Fig{urms} for the turbulent kinetic energy).
As expected, the turbulent magnetic field increases with $\Rey$ for the M
and S runs. Weak super-equipartition fields are observed only for Run
4M in the middle of the convection zone. Runs 4M2, 4S, 3M, and 2M
reach values above $\brmsp/\Beq=0.9$ but do not exceed unity in the
entire convection zone.
The pure SSD (S) runs consistently
exhibit weaker $\brmsp$ compared to their M counterparts.

Interestingly, the field increases with $\Rey\,(=\!\Rm)$ for the M runs, mostly
in the lower half of the convection zone, and appears mostly
independent of $\Rm$ in the upper part of the domain. Only the
highest $\Rm$ run (4M) shows a slightly enhanced field in the upper
part of the domain. In contrast, the S runs show an increase
of $\brmsp$ with $\Rm$ throughout the domain. We interpret this as follows: In the M
runs, where the SSD is still weak, $\brmsp$ is primarily generated by
the tangling of the large-scale magnetic field. This process
seems independent of
$\Rey$ for the upper part of the domain while becoming more effective in
the lower part of the domain as $\Rey$ increases. As the SSD increases
similarly at all radii, it also enhances $\brmsp$ for the highest
$\Rey$ runs, where the SSD field becomes comparable
to the tangled one.

In \Fig{spectra}, we present the spherical-harmonic spectra of
kinematic and magnetic energies along with their ratio for all runs, calculated from near the surface
($r=r_s\equiv0.98\,R$) $\theta\phi$-slices.
The spectral energies $\tilde{E}_{\rm kin} (l)$ and $\tilde{E}_{\rm
  mag} (l)$ are calculated using the following definitions:
\begin{equation}
\sum_l \tilde{E}_{\rm kin} (l) = r_{s} \left. E_{\rm kin}\right
|_{r=r_s}\quad\sum_l \tilde{E}_{\rm mag} (l) = r_{s} \left. E_{\rm mag}\right |_{r=r_s},
\end{equation}
where $l$ is the spherical-harmonic degree,
and the  energy densities are given as
\begin{equation}
\left . E_{\rm kin}\right |_{r=r_s}=\half\left
  . \bra{\rho\uu^{\prime
      2}}_{\theta\phi}\right|_{r=r_s}\quad \left . E_{\rm mag}\right |_{r=r_s}=\half\left
  . \bra{\bb^{\prime 2}}_{\theta\phi}\right|_{r=r_s}.
\end{equation}
We have removed the mean-field contribution ($m=0$) and summed over all other
$m$. Because of our ``wedge'' approach, $l=4$ is the lowest possible $l$.
See \cite{VWKKOCLB17} and \cite{KVKBS19} for details on how to compute spherical
harmonic decompositions from simulations of the presented type.

For low $l$, the kinetic energy spectra (\Fig{spectra}a)
are similar for all runs, with only the hydro runs having slightly higher
energies. At the highest resolutions and $l>50$, the velocity
develops an inertial range, which can be best described by a power-law
of $l^{-11/5}$, illustrated by the compensated spectrum in the inset.
Such a power-law has been predicted by \cite{B59} and \cite{O59},
and also found in \cite{K2021}.
However, as discussed in several studies \citep[e.g.,][]{B92,RR10,XH22,SMP23},
this scaling was obtained for stably stratified media but not for
rotating convection.
Also, the Bogliani-Obukhov scaling should appear only for those larger
scales, which are influenced by the gravity-induced anisotropy,
while being followed by a standard Kolmogorov scaling for smaller scales.
Furthermore, we observe an extended inertial range only at
the highest resolutions, as viscous diffusion affects all
$l$ otherwise.

For the magnetic spectra (\Fig{spectra}b),
 the energies are at low $l$ very similar for all M runs;
only 4M has lower power at $l<10$, which might be caused by some
data loss.\footnote{Unfortunately, we lost the data slices of Run 4M,
  so we used some early slices from Run 4M2, where the mean-field had
  been removed, but not all large-scale fields had yet decayed.}
We do not observe an inertial range with a clear power law as in the H runs.
For the S runs, the energies are lower at all scales compared to the
corresponding M runs.
Yet, Run 4S has higher power than the other S runs, in particular at low
$l$, while seeming indistinguishable from Run 4M for $l>100$.
The field in the S runs being lower than in the M runs, in particular at
low l, is consistent with
nearly all the field at $l>500$ being due to the SSD, whereas tangling
is dominant at $l<500$ for the highest $\Rm$.
Surprisingly, we find no significant difference between the shape of
the spectra of the M and S runs, implying that the spectral properties
of LSD- and SSD-generated fields are very similar,
at least for $l\ge 4$.
We only observe that amplitude differences are more pronounced at $l<100$.

To investigate whether the magnetic field reaches super-equipartition
at certain scales, we also examine the ratio of the magnetic and
kinetic spectra near the surface, as shown in \Fig{spectra}c. Only
in Run~3S, super-equipartition is achieved (around $l\sim1000$), whereas the
corresponding M run has a maximum ratio of 0.6. The lowest diffusivity
runs only achieve just equipartition for $600 < l < 2000$.
It should be stressed
that these spectra are taken close to the surface, where
the horizontally averaged magnetic field is well below equipartition,
as seen in \Fig{brms}. We would expect the 
spectral ratio at larger depths to be higher.
The fanning of the spectra, particularly at high $l$
in the kinetic ones, 
is related to inaccuracies resulting from employing spherical
harmonic decomposition on a spherical grid with incomplete $\theta$ range.

To show more clearly the $\Rey$ dependence of the spectral energy, we plot for
$l=10$ and $100$ the kinetic and magnetic energy as function of $\Rey$
for all runs, see \Fig{spectra}d.
We find that the kinetic energy for $l=10$ seems to decrease mildly with
increasing $\Rey$, interestingly not only for M or S
runs, but also for H runs, hence this cannot be due to a suppression
by the magnetic field as claimed in HK21.
The magnetic energy at $l=10$ of the M runs seems to be independent of $\Rey$, except
a mild decrease for the highest $\Rey$. Only for the S runs, we
see a strong increase.
For the kinetic and magnetic energy
at $l=100$, we find a continuous increase.
Only the kinetic energy seems to saturate for the highest
$\Rey$.
For this scale, we do not find that the strong SSD in the highest
$\Rey$ runs has a marked
influence on the kinetic energies.

If we compare our results with the spectra provided by HK21 and
HKS22, we find two main differences:
The authors observe super-equipartition fields at small scales ($l>100$)
even for their lowest resolution (similar to our Set 2). Additionally,
in their work the kinetic energy in large scales becomes suppressed
at high $\Rey$ because of, as they interpret it, their very strong SSD.
We observe a small decrease in the large-scale energy, but this is
also evident in the hydrodynamic runs.
It is important to note that in HK21 and HKS22, only one pure
hydrodynamic run is studied; therefore there is no $\Rey$
dependence in their case. Furthermore, their simulations only develop
an SSD but no LSD; hence, the influence of a
large-scale field on the dynamics was not studied.
The differences in the spectral properties might 
also be related to these two main
distinctions in the setup as discussed above.

 \begin{figure}[t!]
\begin{center}
\raggedright
\includegraphics[width=0.094\textwidth]{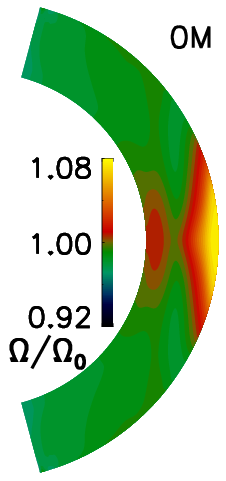}
\includegraphics[width=0.094\textwidth]{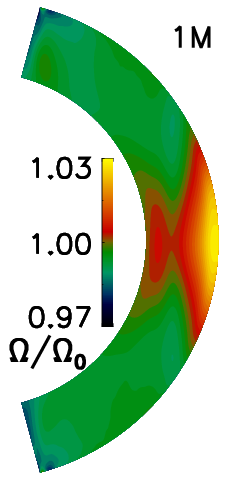}
\includegraphics[width=0.094\textwidth]{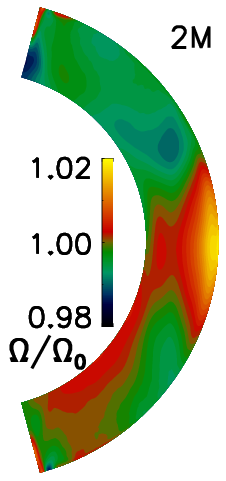}
\includegraphics[width=0.094\textwidth]{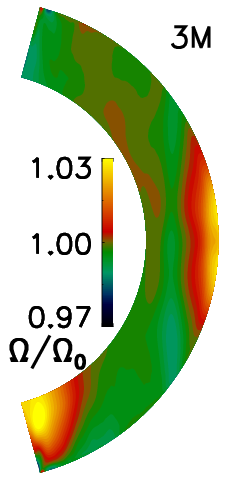}
\includegraphics[width=0.094\textwidth]{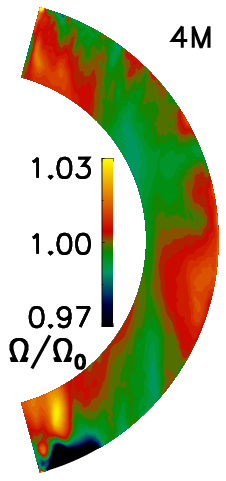}\\
\hfill
\includegraphics[width=0.094\textwidth]{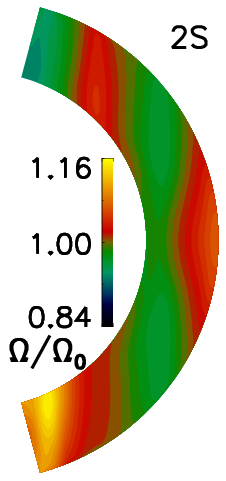}
\includegraphics[width=0.094\textwidth]{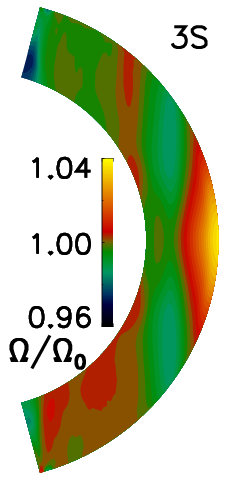}
\includegraphics[width=0.094\textwidth]{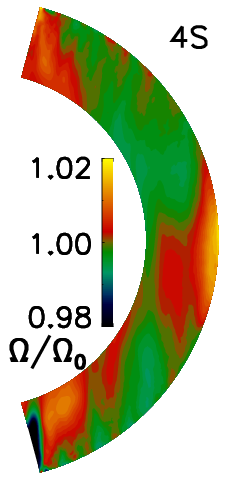}\\
\includegraphics[width=0.094\textwidth]{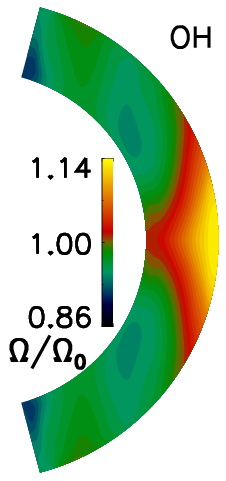}
\includegraphics[width=0.094\textwidth]{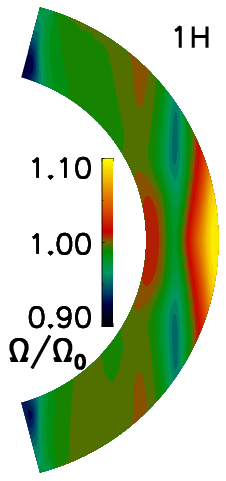}
\includegraphics[width=0.094\textwidth]{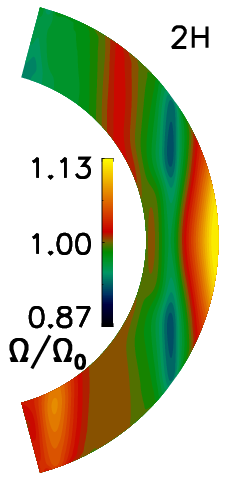}
\includegraphics[width=0.094\textwidth]{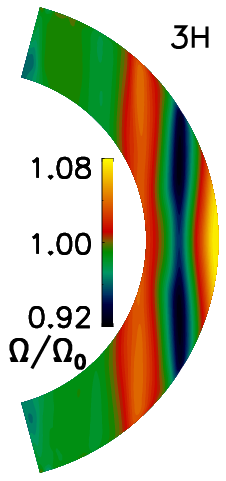}\\
\end{center}\caption{
Differential rotation $\Omega/\Omega_0$, with
$\Omega=\Omega_0 + \mean{u}_\phi/r\sin\theta$ for all runs.
}\label{diffrot}
\end{figure}

\begin{figure}[t!]
\begin{center}
\includegraphics[width=\columnwidth]{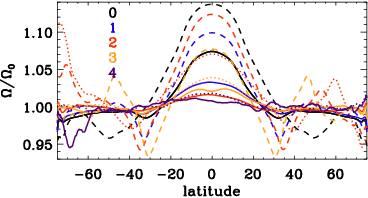}
\includegraphics[width=\columnwidth]{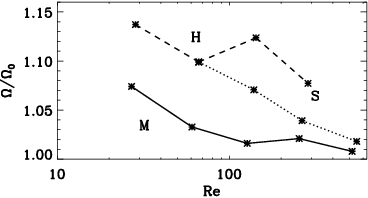}
\end{center}\caption{
Differential rotation of all runs near the surface ($r=0.98\, R$).
Top:  $\Omega/\Omega_0$
as function of latitude. As before, M runs - solid, S runs -
dotted, H runs - dashed lines, and colors indicate the run
sets with different $\Rey$.
Bottom: $\Omega/\Omega_0$ at the equator ($\theta=\pi/2$) as function of $\Rey$.
}\label{omth}
\end{figure}

\begin{figure}[t!]
 \begin{center}
   \includegraphics[width=0.497\textwidth]{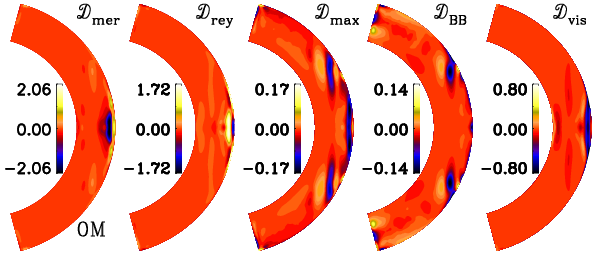}
    \includegraphics[width=0.497\textwidth]{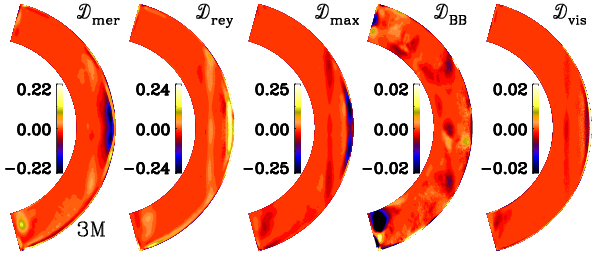}
 \end{center}\caption{
Time-averaged contributions to the
differential rotation energy evolution
for Runs~0M and 3M, as given in \Eqss{eq:Diff_en_u}{eq:Diff_en_v}. The contributions are shown in
units of J/m$^3$s.
}\label{difforce}
\end{figure}

\begin{figure*}[t!]
 \begin{center}
 \includegraphics[width=0.9\textwidth]{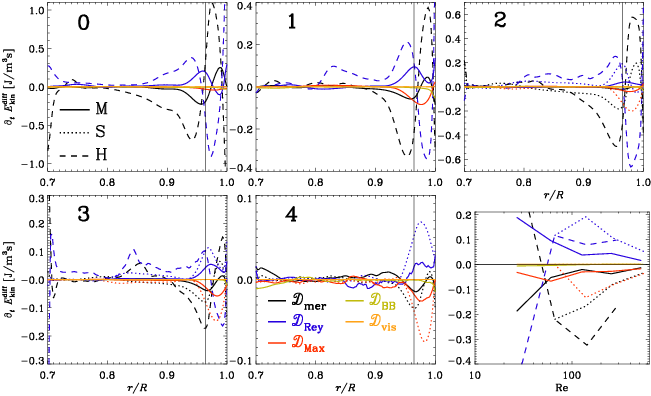}
 \end{center}\caption{
Contributions
to differential rotation energy evolution
 averaged in time and
latitude for all run sets as functions of $r/R$,
see
\Eqss{eq:Diff_en_u}{eq:Diff_en_v}.
The rightmost bottom
panel  shows the contributions as functions of $\Rey$
at $r/R$=0.96
(vertical line in other panels).
}\label{difforce_all}
\end{figure*}

\subsection{Differential rotation and its generators}
\label{sec:diff}

Next, we examine the profiles of differential rotation
$\Omega=\Omega_0 + \mean{u}_\phi/r\sin\theta$
as shown in
\Figs{diffrot}{omth}, and investigate the influence of the magnetic fields
on their generation terms, see
\Figs{difforce}{difforce_all}. As observed from the energy analysis,
the differential rotation is most pronounced in the H runs and
suppressed in the M and S runs.

In the H runs, the contours of constant angular velocity tend to
become more cylindrical towards the more
turbulent regime, while the
maximum $\Omega$
values remain roughly constant across Runs 0H, 1H, and
2H. However, for the highest $\Rey$ (Run 3H),
the differential rotation is
slightly weaker compared to Run 1H. With increasing $\Rey$, more jets
of opposite signs become visible.
Surprisingly, in Run~0H the contours are far less cylindrical compared
to the H runs with higher $\Rey$. We attribute this to the absence of
Busse  columns (or banana cells)  at the lowest $\Rey$, as illustrated in
\Fig{ban_spec}.

As previously noted, the M runs
consistently exhibit weaker differential rotation, with their $\Omega$
profiles being strongly influenced by the magnetic field. We observe a
tendency for the isorotation contours to become less cylindrical with
higher $\Rey$. Excluding some local minima and maxima near the poles,
the differential rotation becomes much weaker in the highly turbulent regime
compared to the less turbulent one. Additionally, in all cases with active
SSD, the differential rotation profile becomes increasingly
hemispherically asymmetric with rising $\Rey$, attributable to the
hemispherically asymmetric magnetic field.
Notably, the
minimum of $\Omega$ at mid-latitudes nearly vanishes in the high-$\Rey$ M runs. 
For the S runs, the profiles at weak SSD (Run 2S)
resemble those of Run 2H. However, Run 3S with stronger SSD exhibits
weaker differential rotation, akin to Run 3M. Comparing the profiles
of Run 4S and 4M reveals hardly any differences. This might be related
to the issues of these runs discussed in \Sec{sec:eng}.

The changes in the
differential profile with $\Rey$ and the presence of the two dynamos
become clearer in the plots of latitudinal distribution and $\Rey$
dependence (\Fig{omth}) of $\Omega$.
It is evident that it significantly decreases near the equator
with increasing $\Rey$ and in the presence of a magnetic
field. Moreover, the jets present in the H runs at mid-latitudes are completely
suppressed in the M runs. Interestingly, the differential rotation at
the equator is already strongly suppressed in Run 2M, where the SSD is
relatively weak, indicating that the suppression is due to the
magnetic field generated by the LSD rather than the
SSD. However, the SSD in Run 4S is capable of suppressing and shaping
the differential rotation similarly to its corresponding Run 4M.

The reduction of shear at high $\Rm$ has also been reported by
\cite{KKOWB17}. HK21 and HKS22 find that at
high $\Rm$, the
differential rotation is also strongly influenced by the magnetic
field, which is consistent with our work. However, in their cases, the
magnetic field is solely due to an SSD, whereas
in our case, the change in differential rotation is mainly due to the
LSD.

Our modeling strategy does not allow us to inspect the actual generation
process of the differential rotation profiles, as we re-start from earlier models.
Hence we can only address reliably the relaxed states.

The differential rotation follows the mean angular momentum balance \citep[see, e.g.][]{R89},
\begin{align}
{\partial_t}\left(s^2\meanrho\Omega\right) =-\nab\cdot
                                              s&\left[s\overline{\rho\uu}\Omega+\overline{\fluc{\left(\rho\uu\right)}\fluc{u}_\phi}
                                              -
                                                 2\nu\meanrho\mean{\SSSS}\cdot\eee_\phi\right.\nonumber  \\
&\left.-\mu_0^{-1}\left(\meanBB\,\mean{B_\phi}+\overline{\fluc{\BB}\fluc{B_\phi}}\right)\right],
\label{eq:angb}
\end{align}
where $s=r\sin\theta$ is the lever arm.
As usual, we neglect the compressible term related to
$\overline{\fluc{\rho}\fluc{\SSSS}}$ in the equation above.

To check which part of the angular momentum flux contributes significantly
to the differential rotation, we calculate the contributions to the
evolution of differential rotation energy density.
This approach has the advantage that
generating contributions are shown as positive values, and destructing ones
as negative.

\begin{equation}
{\partial_t} E^{\rm diff}_{\rm kin} =
(\mean{u_\phi}/s)\partial_t\left(s^2\meanrho\Omega\right) = \half \partial_t (\meanrho \,\mean{u_\phi}^2) = \cal{D}\,.
\label{eq:ang_eng}
\end{equation}
Using the right hand side of \Eq{eq:angb}, we can divide $\cal{D}$
into five different contributions
\begin{alignat} {3}
\cal{D}_{\rm Mer} &=&-&\,(\mean{u_\phi}/s)\, \nab\cdot\left(s^2\overline{\rho\uu}\Omega\right)&\;\;\;&\text{meridional circulation,}  \hspace{-2mm} \label{eq:Diff_en_u}\\
\cal{D}_{\rm Rey}  &=&-&\,(\mean{u_\phi}/s)\, \nab\cdot\left(s\overline{\fluc{\left(\rho\uu\right)}\fluc{u_\phi}}\right)&\;&\text{Reynolds stress,} \label{eq:Diff_en_r}\\
\cal{D}_{\rm Max}
                  &=&&\,(\mean{u_\phi}/s)\,\nab\cdot\left(s\mu_0^{-1}\overline{\fluc{\BB}\fluc{B_\phi}}\right)&\;
                      &\text{SS Maxwell stress,} \label{eq:Diff_en_m}\\
\cal{D}_{\rm BB}
                  &=&&\,(\mean{u_\phi}/s)\,\nab\cdot\left(s\mu_0^{-1}\meanBB\,\mean{B_\phi}\right)&\;&\text{LS Maxwell stress,}\\
\cal{D}_{\rm
  vis}&=&&\,(\mean{u_\phi}/s)\,\nab\cdot\left(2s\nu\meanrho\mean{\SSSS}\cdot\eee_\phi\right)&\;&\text{viscous stress.} \label{eq:Diff_en_v}
\end{alignat}
Due to our choice of azimuthal  averages written out during run time, we
calculate $\cal{D}_{\rm Mer}$ and $\cal{D}_{\rm Rey}$ slightly
differently. However, their sum will be the same. For $\cal{D}_{\rm
  Mer}$, we use
\begin{equation}
\nab\cdot\left(s^2\overline{\rho\uu}\Omega\right)\approx
\nab\cdot\left(s^2\meanrho\,\meanuu\,\Omega\right) + \nab\cdot\left(s^2 \overline{\fluc{\rho}\fluc{\uu}}\Omega_0\right),
\end{equation}
where the second term is calculated by
$-s^2\Omega_0\nab\cdot\left(\meanrho\,\meanuu\right)$ using the
conservation of mass flux
\begin{equation}
\nab\cdot\left({\rho\uu}\right)=
\nab\cdot\left(\overline{\fluc{\rho}\fluc{\uu}}\right) + \nab\cdot\left(\meanrho\,\meanuu\right) =0 \,.
\end{equation}
For $\cal{D}_{\rm Rey}$ we use
\begin{equation}
\nab\cdot\left(s\overline{\fluc{\left(\rho\uu\right)}\fluc{u_\phi}}\right)\approx
\nab\cdot\left(s\overline{\rho\uu u_\phi}\right) - \nab\cdot\left(s\meanrho\,\meanuu\,\meanU_\phi\right).
\end{equation}
Our choice at the end means that the term $\nab\cdot\left(s\,\overline{\fluc{\rho}\fluc{\uu}}\,\meanU_\phi\right)$ is included in
$\cal{D}_{\rm Rey}$ instead of in $\cal{D}_{\rm Mer}$, hence their sum
is the same as in the definitions \Eqs{eq:Diff_en_u}{eq:Diff_en_r}.

We show the time-averaged contributions
exemplarily for Runs 0M and 3M as a meridional plot in \Fig{difforce}
and for all runs as a line plot from additional averaging
over latitude in \Fig{difforce_all}.
For all low $\Rey$ runs, the main balance is between the contributions
of meridional circulation $\cal{D}_{\rm
  mer}$ and Reynolds stresses $\cal{D}_{\rm Rey}$ with a small contribution of
$\cal{D}_{\rm vis}$.
For most of the domain, $\cal{D}_{\rm Rey}$
is generative whereas $\cal{D}_{\rm mer}$ is destructive.
At high $\Rey$, this balance still holds for the H runs, where $\cal{D}_{\rm vis}$
becomes increasingly smaller.
For the M runs, we find two main channels
through which the magnetic field
influences the angular momentum.
On one hand, the  magnetic field suppresses $\cal{D}_{\rm
  mer}$ and $\cal{D}_{\rm Rey}$ to much lower levels, accompanied by some
changes in the spatial distribution.
On the other hand, the contribution from small-scale Maxwell stresses $\cal{D}_{\rm Max}$
becomes comparable to $\cal{D}_{\rm Rey}$, but has mostly negative
values, hence compensating $\cal{D}_{\rm Rey}$.
Surprisingly, this happens already at $\Rey\sim60$, where no SSD is
present.
The direct influence of the large-scale magnetic field ($\cal{D}_{\rm BB}$) 
appears to be significant only near the  surface at high
$\Rey$, where it has a destructive
effect. For the S runs, we find that the contribution of
$\cal{D}_{\rm Rey}$ is not as effectively quenched as in the
corresponding M runs. The magnetic influence primarily comes through
the $\cal{D}_{\rm Max}$ contribution, which has a destructive  effect
across most of the convection zone. Together with the destructive
contribution of $\cal{D}_{\rm mer}$, it balances $\cal{D}_{\rm Rey}$.

From this differential rotation energy balance, we can deduce the
reason why the differential rotation is mostly quenched by the
presence of the LSD rather than the SSD
alone. The quenching of $\cal{D}_{\rm Rey}$ appears to play an
important role here. Only the large-scale field can effectively quench
this contribution, preventing the development of strong differential
rotation. Additionally, the small-scale magnetic field, whether
generated by tangling or by an SSD, creates a
destructive term, $\cal{D}_{\rm Max}$, which further suppresses
the generation of differential rotation.

In theory, Maxwell stresses could also act similarly to Reynolds
stresses in generating differential rotation; however, we find
that their contribution is always destructive.
Previous studies by
\cite{KKOWB17} have shown that the contribution of the Reynolds stress is
balanced by the contribution of meridional circulation, and that this
balance shifts to a balance of Reynolds and Maxwell stress
contributions. This finding has been confirmed by HK21 and
HKS22. However, the latter authors mostly attribute this change in balance
to the presence of an SSD, whereas we find that
this is already the case at moderate $\Rm$, where no SSD is present.

The finding of large-scale magnetic fields suppressing differential
rotation is a result dating back to early magnetoconvection models
\citep[e.g.][]{G83}, but also to many theoretical calculations. Quenching
of the $\Lambda$ effect by magnetism has been intensively studied in
mean-field models \citep[e.g][]{KRK1994,K16,P17} 
and also confirmed by numerical
simulations \citep{KKT04,K19b}.
However, the magnetic field will also affect
the turbulent viscosity in such mean-field models \citep[e.g.][]{KPR94},
therefore differential rotation can also be enhanced \citep{K16}.
Quenching of the $\Lambda$ effect by SSD has been found to be
milder than by a corresponding large-scale magnetic field with the
same strength using simplified numerical simulations \citep{K19c}.

The importance of Busse columns (or banana cells) for the generation of differential
rotation has previously been  pointed out by other authors
\citep[e.g][]{HRY15,FH16,MHT20,BCG22,PK23}.
One of their interpretations is that the prominent presence of
these large-scale convective structures in simulations
makes the differential rotation of the Sun difficult to reproduce. Our results
are in line with this interpretation, because the banana cells seem
to make the differential rotation profile more cylindrical, whereas the
contours of the Sun's differential rotation are radial
over a considerable range of mid-latitudes
\citep[e.g.][]{Schouea98}.

\begin{figure*}[t!]
\begin{tabular}{@{\hspace{0cm}}ccc}
  $B_\phi(r=0.98\, R)$  & $B_\phi(r=0.71\, R)$  & $B_\phi(\theta=25^\circ)$  \\[1mm]
\includegraphics[width=0.32\textwidth]{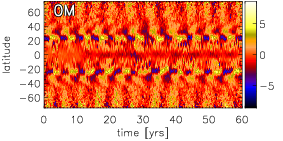}    &
\includegraphics[width=0.32\textwidth]{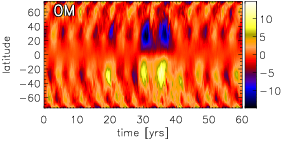}  &
\includegraphics[width=0.32\textwidth]{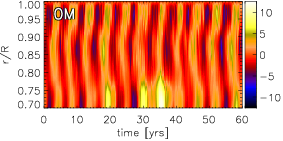} \\
\includegraphics[width=0.32\textwidth]{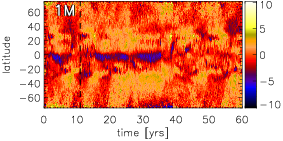} &
\includegraphics[width=0.32\textwidth]{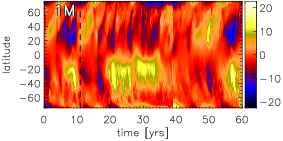} &
\includegraphics[width=0.32\textwidth]{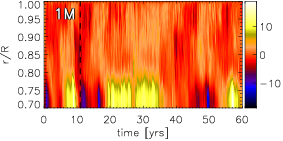} \\
\includegraphics[width=0.32\textwidth]{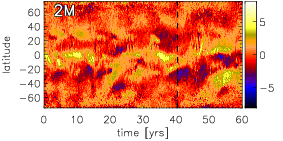} &
\includegraphics[width=0.32\textwidth]{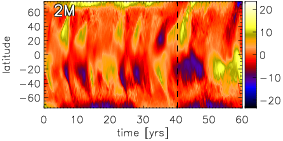} &
\includegraphics[width=0.32\textwidth]{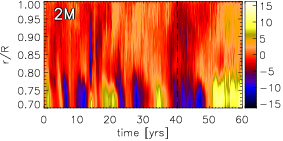} \\
\includegraphics[width=0.32\textwidth]{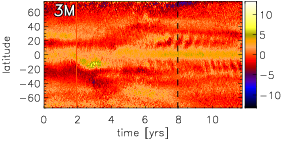} &
\includegraphics[width=0.32\textwidth]{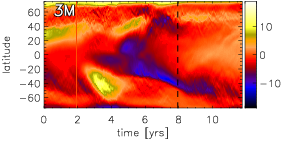} &
\includegraphics[width=0.32\textwidth]{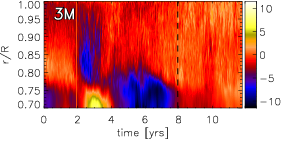}
\end{tabular}
\caption{
Time evolution of the mean toroidal
magnetic field $\mean{B}_\phi$ for all M runs
(except 4M and 4M2). We show time-latitude (butterfly) diagrams at
$r=0.98\, R$ (first column), at $r=0.71\, R$ (second column) and the
time-radius diagram at 25$^\circ$ latitude (north). We note that the time scale for the first three rows are
the same, but different for the last one.
The vertical dashed lines indicate the starting point from which we
use a time interval to compute time averages utilized in the analysis
throughout this paper.
Vertical stripes in Run 3M are due to data loss.
The mean field is in units of kG.
}\label{but}
\end{figure*}

\begin{figure}[t!]
\begin{center}
\includegraphics[width=\columnwidth]{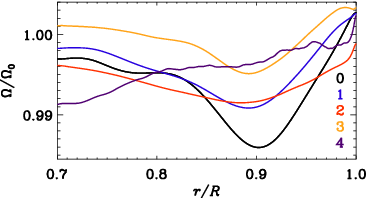}
\end{center}\caption{
Radial profile of the differential rotation $\Omega/\Omega_0$ of all M runs.
We show the radial profile at 25$^\circ$ latitude, which
coincides with the local minimum of $\Omega$ causing the equatorward
migration magnetic field pattern in 0M.
}\label{omr}
\end{figure}

\begin{figure*}[t!]
\includegraphics[width=0.107\textwidth]{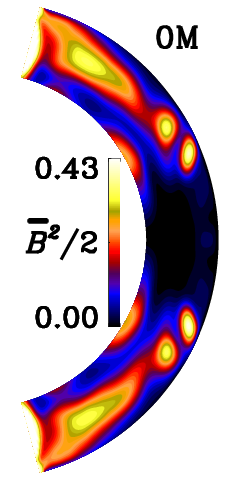}
\includegraphics[width=0.107\textwidth]{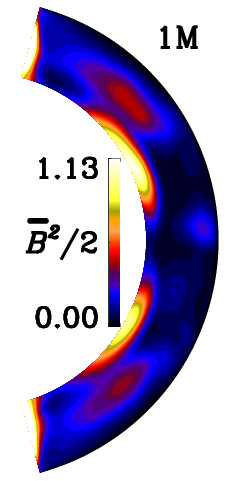}
\includegraphics[width=0.107\textwidth]{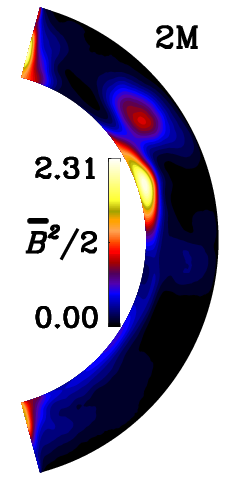}
\hspace{0.107\textwidth}
\includegraphics[width=0.107\textwidth]{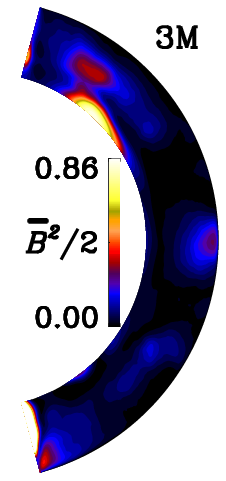}
\hspace{0.107\textwidth}
\includegraphics[width=0.107\textwidth]{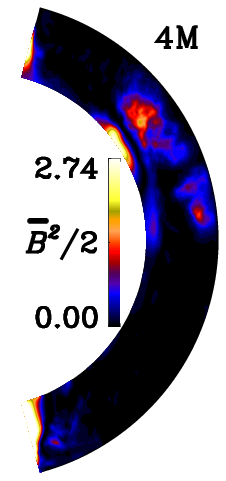}
\includegraphics[width=0.107\textwidth]{Emag_LS_A4}
\hspace{0.107\textwidth}\\
\includegraphics[width=0.107\textwidth]{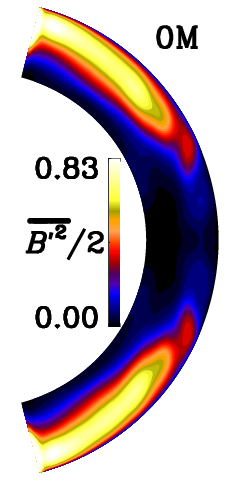}
\includegraphics[width=0.107\textwidth]{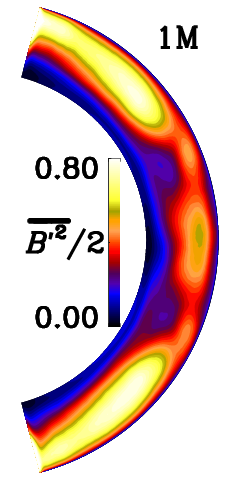}
\includegraphics[width=0.107\textwidth]{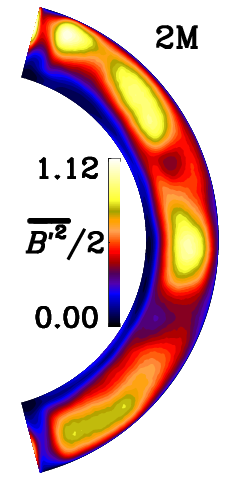}
\includegraphics[width=0.107\textwidth]{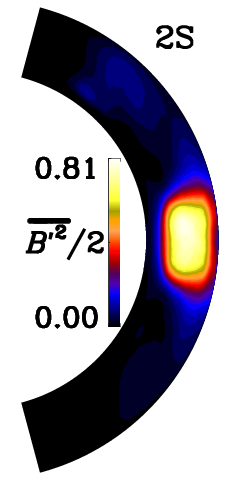}
\includegraphics[width=0.107\textwidth]{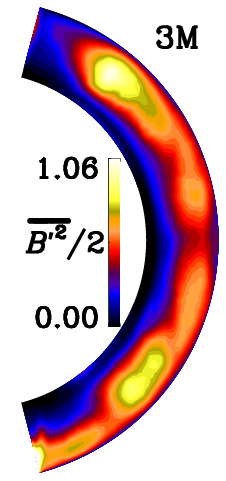}
\includegraphics[width=0.107\textwidth]{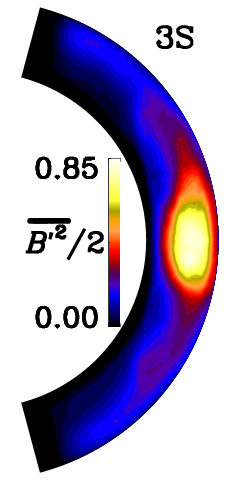}
\includegraphics[width=0.107\textwidth]{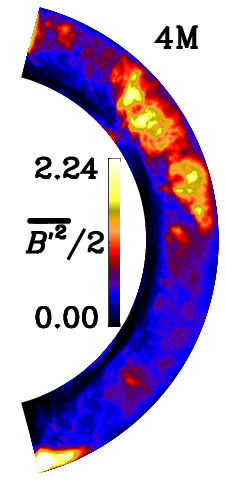}
\includegraphics[width=0.107\textwidth]{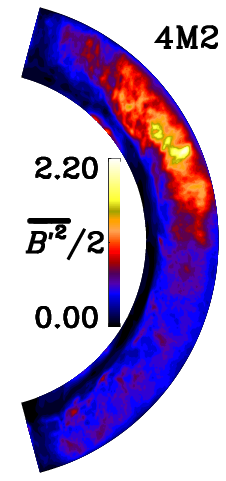}
\includegraphics[width=0.107\textwidth]{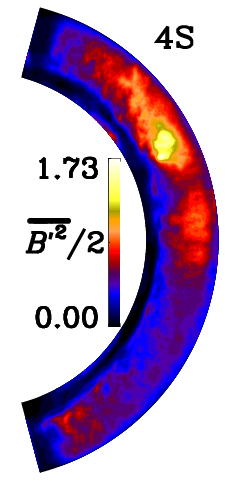}
\caption{
Time-averaged mean magnetic energy of the mean fields $\meanBB^2/2$ (top
row) and fluctuating fields $\overline{\BB^{\prime 2}}$ (bottom) for all M and S
runs in the units of 10$^5$ J/m$^3$ for all magnetic runs (M+S). We disregard 7$^\circ$ near the latitudinal boundary for determining
the minimum and maximum values of the color range.
}\label{Emag_mer}
\end{figure*}

\begin{figure*}[t!]
\begin{center}
  \includegraphics[width=0.107\textwidth]{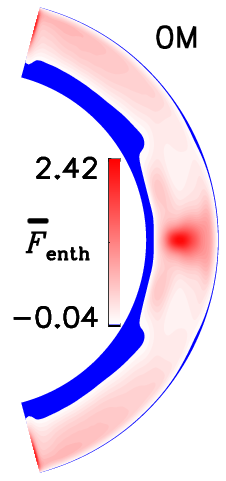}
  \includegraphics[width=0.107\textwidth]{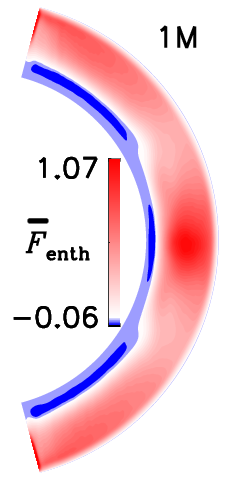}
  \includegraphics[width=0.107\textwidth]{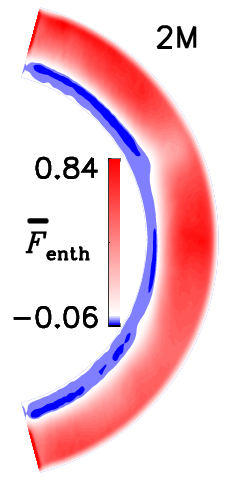}
  \includegraphics[width=0.107\textwidth]{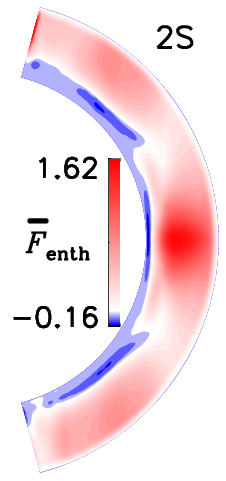}
  \includegraphics[width=0.107\textwidth]{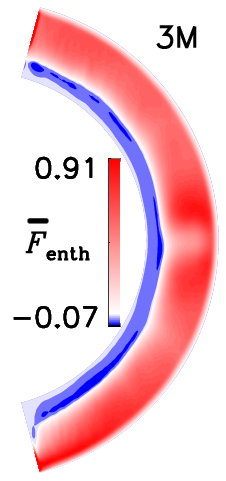}
  \includegraphics[width=0.107\textwidth]{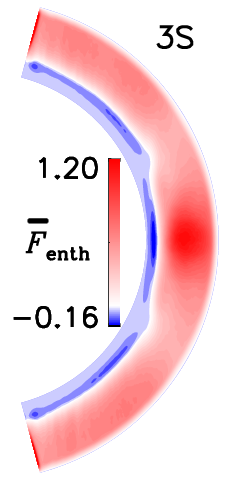}
  \includegraphics[width=0.107\textwidth]{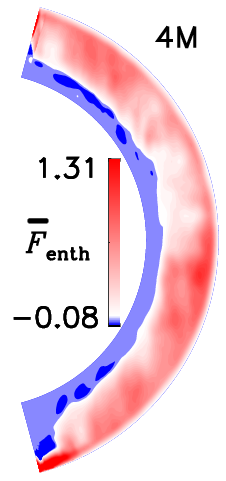}
  \includegraphics[width=0.107\textwidth]{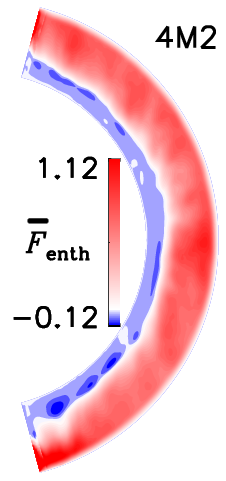}
   \includegraphics[width=0.107\textwidth]{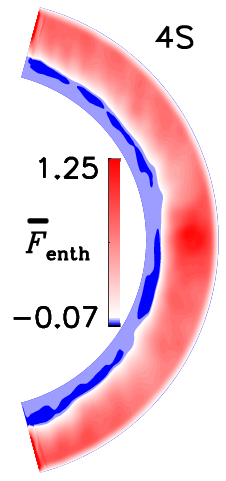}
 \end{center}\caption{
Time-averaged radial enthalpy flux $\mean{F}_{\rm enth}$ normalized by
the input flux $\mean{F}_{\rm tot}$ for all magnetic runs (M+S). We neglect
  7$^\circ$ near the latitudinal boundary for determining the minimum
  and maximum values of the color range.
}\label{Fconv_mer}
\end{figure*}

\subsection{Magnetic field generation}
\label{sec:magfield}

Next, we look at the evolution of the mean toroidal magnetic field $\meanB_\phi$, plotting
in \Fig{but} the butterfly diagrams of Runs 0M, 1M, 2M and 3M; for Run
4M, the time series is too short to form a meaningful diagram.
Run 0M shows a very similar mean field evolution as the run of
\cite{KKOBWKP16}: a regular cycle with equatorward migration
throughout the convection zone, a fast cycle with poleward migration
near the surface close to the equator and a long cycle most pronounced
at the bottom of the domain, capable of
disturbing the other cycles.
This is expected as 0M has the same parameters as the run of
\cite{KKOBWKP16}, except for the use of a Kramers-based
heat conduction instead of a prescribed conductivity profile.
The dynamo mode with equatorward migration, first reported
in \cite{KMB11} and further discussed in
\cite{KMCWB13}, can be clearly explained by an $\alpha\Omega$
Parker dynamo wave \citep[e.g.][]{WKKB14,WRTKKB17}. However, to
obtain the exact period, many other turbulent transport coefficients play a role
\citep{WRVFTK21}. The fast poleward dynamo mode could be identified as
being of $\alpha^2$ type \citep{KKOBWKP16, WRVFTK21}, while the
type of the long-period mode is currently not clear \citep{KKOBWKP16,GKW17}.

Increasing now $\Rey$ and $\Rm$ influences the dynamo solutions: The
clear equatorward migration vanishes for all runs starting from
1M. This is most likely due to changes in the differential rotation profile,
see discussion below. The two other modes, however, still exist in the higher $\Rm$
regime. For the highest values of $\Rm$ (4M), the time series is
unfortunately too short to identify the dynamo cycles safely.
The fast dynamo mode is clearly visible also in the 1M, 2M and even
the 3M run.
That this mode is still visible also in
Run 3M assures
our interpretation as an $\alpha^2$ type dynamo because, as
discussed in \Sec{sec:diff}, the differential rotation becomes very
weak at these high $\Rm$.
The long-cycle dynamo mode is rather irregular and
mostly pronounced in field strength, but it develops also polarity reversals.
Comparing the butterfly diagrams of 1M and 2M, we find that they look
very similar, in particular near the bottom of the domain and at
higher latitudes.
In summary, increasing the Reynolds numbers does not strongly affect
the short and long cycles, still generating significant mean magnetic fields,
but causes the equatorward migrating medium-length mode to vanish.

To investigate why this mode
 vanishes, we inspect in more detail the changes in the differential rotation
profile roughly at the latitudinal and radial locations of
generation of the
previously found equatorward migrating dynamo mode, see \Fig{omr}.
The minimum of $\Omega$ at $r=0.9$ and 25$^\circ$ latitude is very
pronounced in the 0M run, in 1M and 2M it is already much
weaker, but in the run with the
highest $\Rm$ (3M) it has vanished completely.
This fits well with our hypothesis that the negative shear
in this region generates the equatorward migrating dynamo mode seen in
0M \citep{WKKB14,WRTKKB17};
accordingly, when the shear is sufficiently weakened,
this dynamo mode vanishes.
To pinpoint which dynamo is actually working in these simulations,
one needs to measure the turbulent transport coefficients as in
\cite{WRTKKB17} and analyze them via a mean-field model as in
\cite{WRVFTK21}.
Such an analysis is currently under development and will be presented
in a possible follow-up study.

To inspect at which locations the LSD and the SSD operate in our runs,
we show in \Fig{Emag_mer} meridional plots of azimuthally averaged
magnetic energy density of the mean and fluctuating fields separately.
We remind the reader that the magnetic fluctuations in Runs 0M and
1M are entirely generated by the tangling of the large-scale field,
because in these runs no SSD is operating.
In Run 0M, the large-scale field is mostly generated at mid-
to high latitudes near the middle of the convection zone,
which is consistent with earlier findings of runs with similar parameters
\citep{WKKB14,WRTKKB17, KKOBWKP16}. There is also a weaker
field near the bottom of the domain, causing the long-term variation seen in
\Fig{but}. The corresponding small-scale field is located also at mid
to high latitudes as one would expect from the large-scale
field. However, near the bottom of the domain no small-scale field is
generated, due to the weakness of the convective motions in
this area, see \Sec{sec:kramers}.
For the M runs with higher $\Rm$, the mean field at mid-latitudes gradually
vanishes and instead becomes dominant near the bottom of the domain.
The small-scale field is for small $\Rm$ mostly located at high
latitudes, but gradually becomes stronger near the equator.
This increase is partly due to tangling as in 1M, but in particular
for higher $\Rm$ due to the SSD operating in this area, as seen from
runs 2S and 3S.
At the locations, where the small-scale field is strong, the
enthalpy flux and the kinetic energy also reach a maximum, as shown in
\Fig{Fconv_mer} for the enthalpy flux $\meanF_{\rm enth}$.
Such a concentration of $\meanF_{\rm enth}$ near the equator was also observed
in \cite{KVKBS19} and \cite{VK21}. We therefore believe that the
maximum of the small-scale field in these areas is due to the enhanced
convective motions. However, this is only true for the S runs. In the
M runs, the situation is slightly different. In Run 2M, where
already an SSD is present, we indeed see a concentration of small-scale
field near the equator, similar to 2S, but also at higher latitudes,
where the large-scale field is strong. The small-scale field seems still to
 be connected with the entropy flux, as the distribution of
$\overline{{\BB'}^{2}}$ resembles the distribution of $\meanF_{\rm enth}$.

For the high-$\Rm$ runs (3M and 4M), the small-scale field
is not concentrated near the equator as one would expect from their
corresponding S runs. However, here too, the distribution of
$\overline{{\BB'}^2}$ somewhat resembles the distribution of
$\meanF_{\rm enth}$, at least in terms of a minimum near the equator.

Furthermore, even though the mean field is very strong near the bottom
of the domain, the small-scale field is nearly zero. This is most
likely due to the lack of strong convective motions there.
We can explain the strong mean-field at the bottom with the
possibility that the field is generated above and then transported down
by turbulent pumping
or diffusion
and finally diffuses slowly into
this nearly turbulent-free zone, where it can survive for a long time,
because of the lack of strong turbulent diffusion.
As another possibility, the field can be generated locally by the very
strong shear flow, which needs only a weak $\alpha$ counterpart to
form a dynamo loop.

In Runs 4M, 4M2, and 4S, the mean field as well as the small-scale field are
mostly concentrated in the northern hemisphere. This is probably due
to the short integration time. However, even in these runs, there is a
band of weak small-scale magnetic field near the bottom of the domain.
Compared to Runs 2S and 3S,
the SSD in 4S has a larger volume filling factor in
the northern hemisphere, spreading to higher latitudes and outside the
tangent cylinder.
To summarize, the SSD is mostly active near the equator in the S runs and
2M, yet not in 3M and 4M. There is similarity between the
$\meanF_{\rm enth}$  and $\overline{{\BB'}^2}$ distributions,
indicating that both SSD and tangling are strong where the
convection is strong.

\begin{figure*}[t!]
\begin{center}
\includegraphics[width=0.12\textwidth]{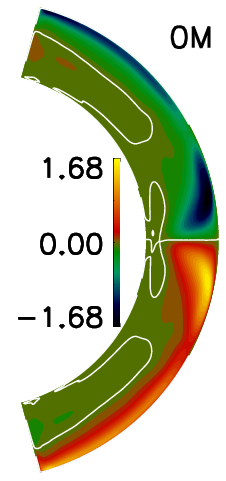}
\includegraphics[width=0.12\textwidth]{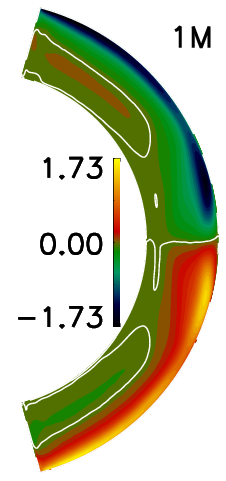}
\includegraphics[width=0.12\textwidth]{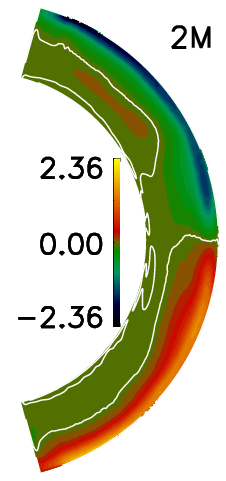}
\includegraphics[width=0.12\textwidth]{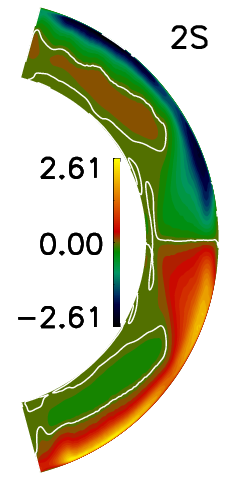}
\includegraphics[width=0.12\textwidth]{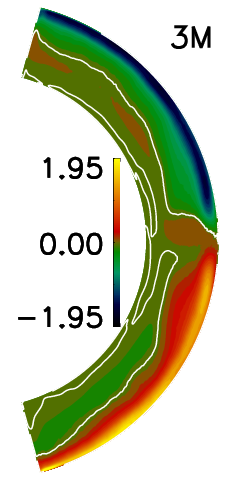}
\includegraphics[width=0.12\textwidth]{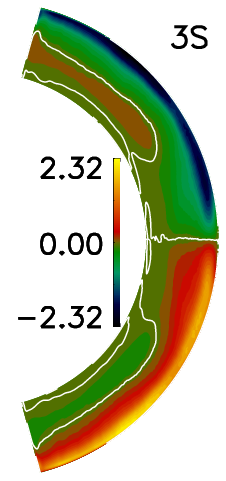}
\includegraphics[width=0.12\textwidth]{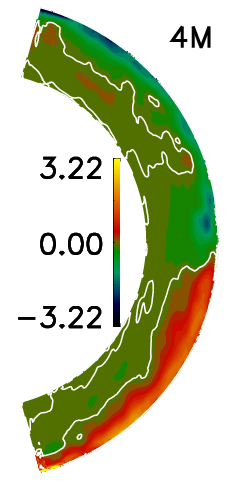}
\includegraphics[width=0.12\textwidth]{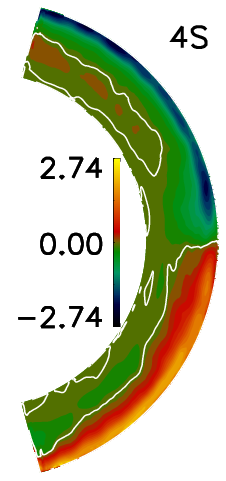}\\
\end{center}\caption{
Time-averaged kinetic helicity $H_{\rm kin}=\overline{\fluc{\oo}\cdot\fluc{\uu}}$ in units of $10^{-3}$m$/$s$^2$
for M and S runs. White contours indicate zero values.
We smooth the data over 100 neighboring points for runs A4M, A4S to
make the data less noisy.
}\label{kinhel}
\end{figure*}

\begin{figure*}[t!]
\begin{center}
\includegraphics[width=0.12\textwidth]{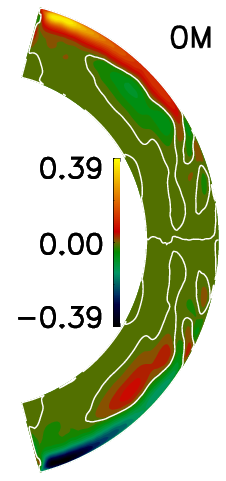}
\includegraphics[width=0.12\textwidth]{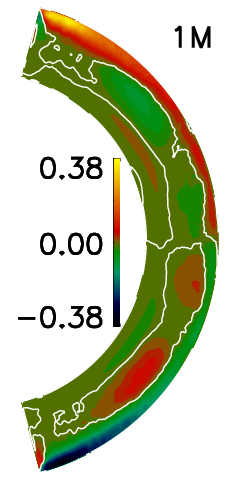}
\includegraphics[width=0.12\textwidth]{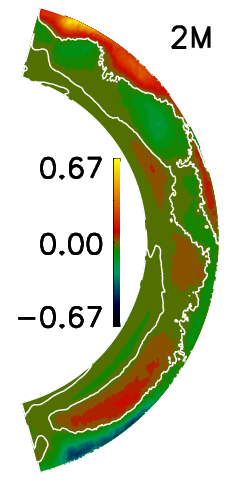}
\includegraphics[width=0.12\textwidth]{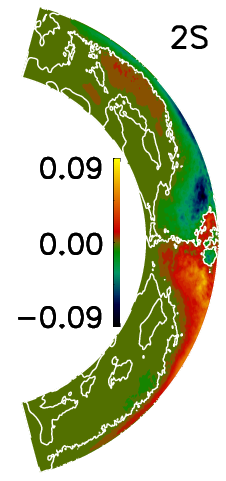}
\includegraphics[width=0.12\textwidth]{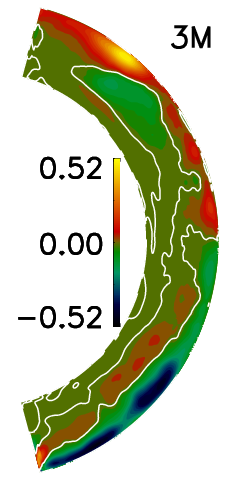}
\includegraphics[width=0.12\textwidth]{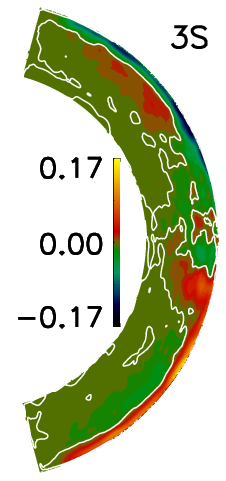}
\includegraphics[width=0.12\textwidth]{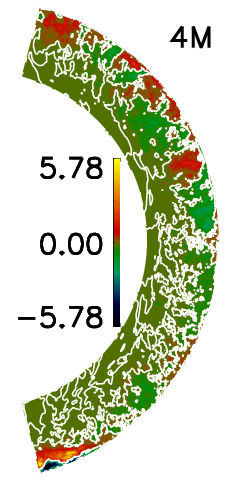}
\includegraphics[width=0.12\textwidth]{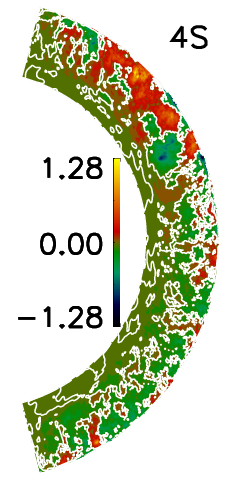}\\
\end{center}\caption{
Time-averaged current helicity  $H_{\rm cur}=\overline{\fluc{\JJ}\cdot\fluc{\BB}}/\meanrho$
  in units of $10^{-3}$m$/$s$^2$ for M and S runs.
 White contours indicate zero values. We smooth the data over 40 and
 100 neighboring points for runs 3M, 3S and 4M, 4S, respectively to
make the data less noisy.
}\label{curhel}
\end{figure*}

\subsection{Kinetic and current helicities}
\label{sec:hel}

One of the most important turbulent dynamo processes is the $\alpha$
effect \citep{SKR66}.
Its strength can be estimated
by the kinetic helicity density $H_{\rm
  kin}=\overline{\fluc{\oo}\cdot\fluc{\uu}}$, and the magnetic
influence on it by the current helicity density $H_{\rm
  curr}=\overline{\fluc{\JJ}\cdot\fluc{\BB}}/\meanrho$ \citep[e.g.][]{PFL76}, here defined
with an $1/\meanrho$ factor:
\begin{equation}
  \alpha\approx-H_{\rm kin}/3\tau_{\rm c} + H_{\rm curr}/3\tau_{\rm c},
  \label{eq:alpha}
\end{equation}
where $\tau_{\rm c}$ is the turbulent correlation time, which in
general does not need to be the same for the first and second term.
In this work, we only look at these proxies (\Figs{kinhel}{curhel})
and leave the detailed analysis of the $\alpha$ tensor and other
turbulent transport coefficients to a future study.

As usual in rotating
convection simulations,
$H_{\rm kin}$ is predominantly negative  in the
upper half of the convection zone and positive below in the northern
hemisphere while having the same pattern but opposite sign in the
southern hemisphere, see \Fig{kinhel}.
We find that the profile of $H_{\rm kin}$ is not much influenced by
the increase of $\Rey$ nor showing a large difference between S and M
runs. We only find that the peak values of $H_{\rm kin}$ show a
tendency to increase with increasing $\Rey$ and that in Run 0M the maxima near
the equator extend further into the convection zone.

The current helicity $H_{\rm curr}$ exhibits a different
behavior. Similar to earlier results \citep[e.g.,][]{WK20}, it is
positive near the surface in the northern hemisphere, negative in the
bulk, and positive again near the bottom of the domain for most of the
M runs while having an analogous profile but with the opposite
sign in the southern hemisphere. Increasing $\Rey$ leaves the profile
mostly unchanged but leads to an increase in the peak values, at least
when comparing 0M and 1M with 2M and 3M.

Interestingly, the current helicity $H_{\rm curr}$, see \Fig{curhel}, from the pure SSD runs follows the
pattern of $H_{\rm kin}$ rather than 
that of $H_{\rm curr}$ of the
corresponding M runs. The fact that the sign of $H_{\rm curr}$ is
opposite in the cases with and without LSD is in line with the results of \cite{WBM12}, where the
authors explain this sign change
from simulations \citep{WBM11} and observations \citep{BSBG11} by a simple model.
It suggests a sign change in $H_{\rm curr}$ 
to be expected when     
an LSD is absent compared to when it is present.

$H_{\rm curr}$ generated by the SSD is clearly smaller than the
one generated by an LSD. Furthermore, $|H_{\rm curr}|$ is always
(except for 4M)
smaller than $|H_{\rm kin}|$ and therefore should not influence the
$\alpha$ effect much, following \Eq{eq:alpha}. However, the work by
\cite{WK20} shows that this approximation does not hold when comparing
with $\alpha$ determined by the test-field method in cases where the
magnitudes of $H_{\rm curr}$ and $H_{\rm kin}$ become comparable.

\section{Conclusions}
\label{sec:conclusions}

We have conducted global convective dynamo simulations of solar-like
stars, wherein we varied viscosity, magnetic diffusivity, and SGS
heat diffusivity to examine how the solutions depend on the fluid and
magnetic Reynolds numbers. This enabled us to investigate the
interaction between SSD and LSD and their effects on the overall
dynamics. As a novel approach, we additionally investigated SSD in
LSD-capable systems in isolation
by suppressing the large-scale magnetic field.

As an outcome of these simulations, we have identified the following
results: Magnetic runs with 
$\Rm\ge140$ 
can excite an SSD, with its
magnetic energy becoming comparable to that of the LSD at
$\Rm\approx550$.
The total magnetic field 
energy appears to reach a maximum
at $\Rm=120$ and decreases for runs with higher $\Rm$.
Although the
SSD becomes stronger at higher $\Rm$, the energy in the fluctuating
field mildly decreases.
As the mean field decreases by
more than half compared to
Runs 2M and 3M, the total magnetic energy is significantly weaker.

The scale of large-scale convective cells, also known as Banana cells, does not depend on
the Reynolds numbers or the presence of LSD and SSD. The depth of the
sub-adiabatic layers is mostly independent of $\Rey$, except at
mid-latitudes. However, the Deardorff layer becomes thinner, allowing
for a thicker overshoot and radiative zone for higher $\Rey$. The turbulent velocity
$\urmsp$ increases with $\Rey$ until $\Rey=120$, after which it
slightly decreases, even in the H runs. This indicates that the
decrease at high $\Rey$ is not due to the presence of SSD or LSD as
found by HK21 and HKS22.
The magnetic field in our simulations does not reach strong
super-equipartition w.r.t. turbulent kinetic energy. Additionally, at
small scales, the magnetic field is mostly at
subequipartition. Furthermore, the energy in the
 large convective scales mildly
decreases with $\Rey$, but this occurs in the H runs as well, suggesting that it
cannot be attributed to the effects of SSD or LSD.

Differential rotation is strongly quenched in M runs with high $\Rey$,
primarily due to the magnetic field of LSD rather than of SSD,
agreeing with many earlier analytical and numerical results
\citep[e.g.][]{G83,KRK1994,KKT04,K16,P17,K19b}.
The magnetic field affects the angular momentum distribution
via the suppression of the Reynolds stresses and the emergence of
strong Maxwell stresses.
The effects of the Maxwell stresses are dominated by the
contributions of the small-scale fields, which are, however, 
 mostly due to tangling of thelarge-scale field and not the SSD.
This contradicts the recent findings of \citep[][]{HK21,HKS22}, who
argued that SSD is the most important driver of fluctuations and,
through them, affects the angular momentum balance.

The evolution of large-scale fields shows only a weak 
dependence on
$\Rm$, with the equatorward migrating field mode disappearing for
$\Rm\geq60$ due to weaker shear at mid-latitudes. The irregular
low-frequency mode at the bottom of the domain persists, and even the
high-frequency mode near the surface is present in all relevant M
runs.
SSD is strongest in areas where the enthalpy flux is maximal,
typically near the equator, where turbulent energy reaches its
peak. The profiles of kinetic and current helicity do not vary much
with $\Rm$, but there is a tendency for a mild increase in their peak values
with $\Rm$. Interestingly, current helicity generated by pure SSD has
the opposite sign to that of LSD.
Our work shows that it is important to study the SSD-LSD interaction
to fully understand the dynamics in the Sun and other stars.

\begin{acknowledgements}
We thank the anonymous referee for 
their very useful comments and
suggestions.
The simulations have been carried out on SuperMUC-NG using the PRACE
project Access Call 20 INTERDYNS project, on the Max Planck
supercomputer at RZG in Garching and in the facilities hosted by the
CSC---IT Center for Science in Espoo, Finland, which are financed by the
Finnish ministry of education.
This project has received funding from the European Research Council
(ERC) under the European Union's Horizon 2020 research and innovation
programme (grant agreement n:os 818665 ``UniSDyn'' and 101101005 ``SYCOS''), and has been
supported from the Academy of Finland Centre of Excellence ReSoLVE
(project number 307411).
This work was done in collaboration with the COFFIES DRIVE Science Center.

\end{acknowledgements}

\bibliographystyle{aa}
\bibliography{paper}

\begin{appendix}

\section{Diffusivity profiles}
\label{sec:diffprof}

\begin{figure}[h!]
\begin{center}
\includegraphics[width=\columnwidth]{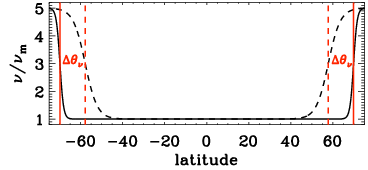}
\end{center}\caption{
Latitudinal diffusivity profiles. Here shown for the viscosity $\nu$
with width $\Delta\theta_\nu=5^\circ$ (solid line) and $17^\circ$
(dashed), also directly indicated by vertical red lines. $\nu_{\rm m}$
is the equatorial value of $\nu$.
}\label{diffprof}
\end{figure}

To avoid numerical artifacts near the latitudinal boundaries, we apply a profile for the diffusivities $\nu$ and $\eta$,
increasing towards these
boundaries. The profiles' shape is shown in \Fig{diffprof}, it is determined by its width
$\Delta\theta_\nu$ and $\Delta\theta_\eta$, respectively, and the ratio between the boundary value and the value at
the equator,
\begin{equation}
\Delta_{\nu\eta}
\equiv{\nu\,(\theta=\Theta_0,\pi-\Theta_0)\over\nu\,(\theta=\pi/2)} =
{\eta\,(\theta=\Theta_0,\pi-\Theta_0)\over\eta\,(\theta=\pi/2)}.
\end{equation}
We choose the minimal
possible values for $\Delta\theta_\nu$, $\Delta\theta_\eta$, and $\Delta_{\nu\eta}$, which keep the runs stable.
It turns out that these adjustments
are needed for runs with $\Rey\ge61$, see \Tab{runs} for details.

Furthermore, for high resolution simulations ($\Rey\ge260$), the runs
have a tendency to produce vortex-like structures at high latitudes. 
They are as such an interesting phenomenon \citep{KMH11,MKH11}, but
they decrease the time step significantly. Hence, we decided to
suppress them in this work by choosing
$\Delta\theta_\nu=17^\circ$ and postpone
their detailed study to the future.

\section{Slope-limited diffusion}
\label{sec:sld}

\begin{figure}[t!]
\begin{center}
\includegraphics[width=\columnwidth]{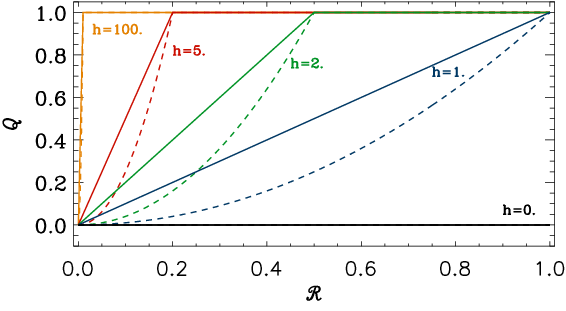}
\end{center}\caption{
Visualization of \Eq{eq:sld_q}: $Q$ as function of the slope ratio
$\cal{R}$ for various $h_{\rm sld}$ (colored lines) with  $n_{\rm sld}$=1 (solid) and
$n_{\rm sld}$=2 (dashed).
}\label{slope}
\end{figure}

Besides the constant diffusivity $\nu$, the H runs need additional
explicit numerical diffusion to be stable. Our choice of using an enhanced luminosity at
the bottom boundary, \citep[see][for details]{KMCWB13}, causes the local
Mach number $\Ma=|\uu|/\cs$ to increase near the surface
to unusually high values.
In the H runs,
this can  lead to numerical instabilities
whereas the M and S runs are stable enough due to the presence of the
magnetic field and hence do not need any
additional diffusion. To let the numerical
diffusion act only in regions, where it is needed, we employ slope-limited
diffusion (SLD), newly implemented to the Pencil Code. It turns out that with our choice of
parameters as specified below, we are able to stabilize the runs without
influencing the overall dynamics much. In the following, we will
briefly describe implementation and parameter choice.
Thereby, we follow roughly \cite{Rempel09} and \cite{Rempel14}.

The main idea is to define the diffusive flux $f$ 
of a velocity component $u$
based on a slope
limiter.
At the cell interface $k+1/2$, it is given by
\begin{equation}
f_{k+1/2} = - \half c^{\rm sld}_{k+1/2}\, Q_{k+1/2}\, (u^{\rm R}_{k+1/2} -u^{\rm L}_{k+1/2}),
\label{diff_flux}
\end{equation}
where the subscript $k$ indicates the cell or grid-point in one particular 
coordinate direction, $u^{R}_{k+1/2}$ and  $u^{L}_{k+1/2}$ are the right and left
values at the cell interface of the velocity component and $c^{\rm sld}$ is the characteristic speed, defined
below.
$u^{\rm R}$ and  $u^{\rm L}$ are defined via
\begin{alignat}{2}
 u^{\rm L}_{k-1/2} &= u_{k-1} &&+ \Delta u_{k-1}\\
 u^{\rm R}_{k-1/2} &= u_{k} &&- \Delta u_{k}\\
 u^{\rm L}_{k+1/2} &= u_{k} &&+ \Delta u_{k}\\
 u^{\rm R}_{k+1/2} &= u_{k+1} &&- \Delta u_{k+1}
\end{alignat}
with the estimated slopes
\begin{equation}
\Delta u_{k} = \text{minmod}\, (u_k - u_{k-1}, u_{k+1} - u_k),
\end{equation}
where the minmod function is defined as
\begin{equation}
\text{minmod}\,(a,b)=\half\sgn(a) \max\left[0,\min\left( |a|,\sgn(a)\, b\right)\right],
\end{equation}
meaning
\begin{eqnarray}
\text{minmod}\,\big(a,b\big) = \left.\begin{cases}
\:+\half \min\big( |a|, |b| \big) \\
\:\phantom{+}0\\
\:\phantom{+}0\\
\:-\half \min\big( |a|, |b| \big)
\end{cases}\hspace{-3mm} \right\}\:\text{for}\;\begin{cases}
\:a>0,\, b>0  \\
\:a<0,\, b>0 \\
\:a>0,\, b<0 \\
\:a<0,\, b<0
\end{cases} \hspace{-3mm} .
\end{eqnarray}
The diffusive flux is additionally
adjusted by
the factor
\begin{equation}
Q_{k+1/2}({\cal R}_{k+1/2}) = \left[\min( 1, h_{\rm sld}\,{\cal
    R}_{k+1/2})\right]^{n_{\rm sld}},
\label{eq:sld_q}
\end{equation}
which controls its non-linearity
by the power $n_{\rm sld}$
and has values between 0 and 1.
The parameter $h_{\rm sld}$ sets the strength of the diffusion for a
given slope. If $h_{\rm sld}$=$\infty$, i.e. $Q=1$, \Eq{diff_flux}
represent a linear 2-order Lax-Friedrichs-scheme. The lower
$h_{\rm sld}$, the less diffusive is the scheme.
The power $n_{\rm sld}$ can reduce the diffusion even
further for small slopes.
The slope ratio ${\cal R}$ is defined as
\begin{equation}
{\cal R}_{k+1/2} = { | u^{R}_{k+1/2} -u^{L}_{k+1/2} | \over  | u_{k+1} -u_{k} |}\,.
\end{equation}
It relates the $u$ difference at the cell
interface to the difference between the cell centers, therefore indicating
the relative slope strength.
Equation~\ref{eq:sld_q} implies that in regions where ${\cal
  R}_{k+1/2} \ge 1/h_{\rm sld}$, the diffusive flux is maximal
and for regions, where ${\cal R}_{k+1/2} < 1/h_{\rm sld}$ the
diffusive flux is reduced, see \Fig{slope} for an illustration.
In this work, we use $h_{\rm sld}=2$ and $n_{\rm sld}=1$ for the
density and $h_{\rm sld}=1$ and $n_{\rm sld}=2$
for all velocity components.
We note here that in \cite{Rempel09} a similar scheme is used, which corresponds
to $h_{\rm sld}=1$ and $n_{\rm sld}=2$. In the work of \cite{Rempel14},
a different expression for $Q({\cal R})$ is used -- see their Eq. (10),
employing only one parameter instead of two. Detailed tests indicate no
significant differences between his and our scheme.

The characteristic speed is
derived from the signal (advection and wave) speeds in the system:
\begin{equation}
c^{\rm sld} = w^{\rm sld}_{\rm hyd} 
|\uu|+
w^{\rm sld}_{\rm sound} \cs + w^{\rm sld}_{\rm mag} v_{\rm A},
\end{equation}
where $v_{\rm A}$ is the Alfv\'{e}n speed, and the
weights
$w^{\rm sld}_\ast$ can be
chosen depending on the nature of the problem. In this work, we set $w^{\rm
  sld}_{\rm hyd}=1$ and $w^{\rm sld}_{\rm sound}=0.001$; there is
no magnetic contribution because we use  SLD
only for purely hydrodynamic runs.
The intercell values of $c^{\rm sld}$ are calculated
by linear interpolation:
\begin{equation}
c^{\rm sld}_{k+1/2} =  {c^{\rm sld}_{k} + c^{\rm sld}_{k+1} \over 2}.
\end{equation}
Connecting the strength of the
SLD term to the 
signal speeds via
$c^{\rm sld}$ makes an extra time step constraint unnecessary.

The calculation of the diffusive fluxes is now performed at each grid
point for all three directions separately. With these three fluxes we
can for a scalar quantity
form a diffusive flux vector ${\ff}_{\rm sld}=(f_q)$, where $q$ indicates
the coordinate direction. If the diffusive fluxes are calculated for 
a vector quantity, then each vector component builds its own flux
vector and these together form a diffusive flux tensor $\FFF_{\rm sld}$.

Finally, the diffusive fluxes are added to the momentum
and continuum equations via
\begin{eqnarray}
{\DD \uu\over\DD t} = ... - \nab^{2\rm nd}\cdot \FFF^u_{\rm sld}, \quad \quad
{\DD \ln\rho\over\DD t} = ... - \nab^{2\rm nd}\cdot {\ff}^\rho_{\rm sld},
\end{eqnarray}
where $\FFF^u_{\rm sld}$ is the SLD tensor for the
velocity and ${\ff}^\rho_{\rm sld}$ is the SLD
vector for the density.
For simplicity, we use 2nd order
finite differences for the
divergence ($\nab^{2\rm nd}$):
\label{divF_cartI}
\begin{equation}
\nab^{2\rm nd}\cdot {\ff}_{\rm sld} = \sum_q\, \partial^{2\rm nd}_q\, f_q= \sum_q{f_q(q_{k_+}) - {f_q(q_{k_-})}\over{q_{k_+}-q_{k_-}}},
\label{divF_cartII}
\end{equation}
where ${\ff}_{\rm sld}$ stands for ${\ff}_{\rm sld}^\rho$ or for one
of the column vectors of $\FFF^u_{\rm sld}$ and $q$ signifies the
coordinate, i.e. $x$, $y$ or $z$.
$q_k$ denotes the grid point under consideration in the
direction of the coordinate $q$ and $q_{k_+}$ and $q_{k_-}$ are short for
$q_{k+1/2}$ and $q_{k-1/2}$, respectively. 
For brevity, we show only that argument of $f_q$, with respect to which the derivation is performed. 
For $\FFF^u_{\rm sld}$, we calculate the divergence 
analogously for each
component.

For spherical coordinates, \Eq{divF_cartII} is modified and reads
for an SLD vector such as $\ff^\rho_{\rm sld}$
\begin{align}
&\nab^{2\rm nd}\!\cdot \:\ff_{\rm sld} 
\nonumber\\
&={ {\rkp^2 f_r(\rkp) - \rkm^2 f_r(\rkm)}\over
    {r^2_k (\rkp-\rkm)}}
+{{\sin(\tkp)f_\theta(\tkp)- \sin(\tkm)f_\theta(\tkm)}\over{r_k\sin(\theta_k) (\tkp-\tkm)}}\nonumber\\
&\phantom{=}+{{f_\phi(\pkp) - f_\phi(\pkm)}\over{r_k\sin(\theta_k) (\pkp-\pkm)}} \\
\end{align}
for an SLD tensor such as ${\FFF}^u_{\rm sld}$
\begin{eqnarray}
[\nab^{2\rm nd}\cdot \FFF_{\rm sld}]_r &=&{\rkp^2 f^r_r(\rkp) - \rkm^2 f^r_r(\rkm)}\over{r^2_k (\rkp-\rkm)}\nonumber\\
&+&{\sin(\tkp)f^r_\theta(\tkp)- \sin(\tkm)f^r_\theta(\tkm)}\over{r_k\sin(\theta_k) (\tkp-\tkm)}\nonumber\\
&+&{f^r_\phi(\pkp) - f^r_\phi(\pkm)}\over{r_k\sin(\theta_k) (\pkp - \pkm)}\\
&-& {{f^\theta_\theta(\tkp) + f^\theta_\theta(\tkm)}\over{2 r_k}}
- {{f^\phi_\phi(\pkp) + f^\phi_\phi(\pkm)}\over {2 r_k}} , \nonumber
\end{eqnarray}
\begin{eqnarray}
[\nab^{2\rm nd}\cdot \FFF_{\rm sld}]_\theta &=&{\rkp^2 f^\theta_r(\rkp) - \rkm^2 f^\theta_r(\rkm)}\over{r_k^2 (\rkp-\rkm)}\nonumber\\
&+&{\sin(\tkp)f^\theta_\theta(\tkp)- \sin(\tkm)f^\theta_\theta(\tkm)}\over{ r_k\sin(\theta_k) (\tkp-\tkm)}\\
&+&{{f^\theta_\phi(\pkp) - f^\theta_\phi(\pkm)}\over{r_k\sin(\theta_k) (\pkp-\pkm)}}
- {{f^r_\theta(\tkp) + f^r_\theta(\tkm)}\over{2 r_k}} \nonumber \\
&-& {{\cot{(\theta_k)}\left(f^\phi_\phi(\pkp) + f^\phi_\phi(\pkm)\right)}\over{2 r_k}} , \nonumber
\end{eqnarray}
\begin{eqnarray}
[\nab^{2\rm nd}\cdot \FFF_{\rm sld}]_\phi &=&{\rkp^2 f^\phi_r(\rkp) - \rkm^2 f^\phi_r(\rkm)}\over{r^2_k(\rkp-\rkm)}\nonumber\\
&+&{\sin(\tkp)f^\phi_\theta(\tkp)- \sin(\tkm)f^\phi_\theta(\tkm)}\over{r_k\sin(\theta_k) (\tkp-\tkm)}\\
&+&{{f^\phi_\phi(\pkp) - f^\phi_\phi(\pkm)}\over{r_k\sin(\theta_k) (\pkp-\pkm)}}
- {{f^r_\phi(\pkp) + f^r_\phi(\pkm)}\over{2 r_k}}\nonumber\\
&-& {{\cot{(\theta_k)}\left(f^\theta_\phi(\pkp) + f^\theta_\phi(\pkm)\right)}\over{2 r_k}},\nonumber
\end{eqnarray}
where the superscripts of $f$ indicate the quantity the
diffusive flux is calculated for, e.g. $r$ for $u_r$.

The viscous heat due to  SLD,
${\cal H}^{\rm sld}$,
is defined at grid point $k$ as
\begin{align}
{\cal H}^{\rm sld} = \half\sum_{p,q} &\left\{  \hspace{4mm} f^{p}_q(q_{k_-}) {\rho(q_k)
                        u_p(q_k) - \rho(q_{k-1}) u_p(q_{k-1})\over q_k-q_{k-1}} \right.\\
                              &\hspace{2mm}+     \left.
                                f^{p}_q(q_{k_+}) {\rho(q_{k+1}) u_p(q_{k+1}) - \rho(q_k) u_p(q_k)\over q_{k+1}-q_k}\:\right \}. \nonumber
\end{align}
where $p,q$ denote the Cartesian coordinates.
 In spherical coordinates, this expression will be
modified taking into account the appropriate derivatives.
We note here that the viscous heating implementation in MURAM
\citep{Rempel09,Rempel14} only takes into account the derivative of
the velocity and not the momentum, i.e. it neglects the changes in the density.

Using a mass diffusion introduces an additional mass flux, for which we
compensated in the momentum and energy equations.

\section{Spectra of banana cells}
\label{sec:ban}

\begin{figure}[t!]
\begin{center}
\includegraphics[width=\columnwidth]{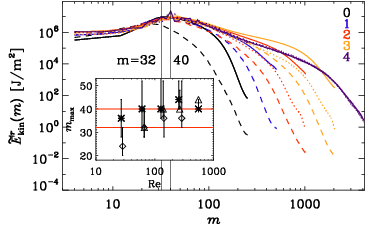}
\end{center}\caption{
Longitudinal power spectra of the radial kinetic energy density, $\widetilde{E}^r_{\rm kin}$, near the surface
$r$=0.98$\,R$ for a narrow latitudinal band around the equator ($\pm\,
7.5\degr$). The colors indicate the run sets and the line style/symbols the
run type: solid/asterisk (M), dashed/diamonds (H) and dotted/triangles
(S).
The spectra are averaged over the $\theta$ bands and time.
The inlay shows the $m$ value of the maxima as a
function of $\Rey$ for all runs.
The errors are calculated using an 80\% range around the peak.
The values $m=32$ and $m=40$ are indicated by black vertical and (in inlay)  red
horizontal lines, respectively.
}\label{ban_spec}
\end{figure}

To investigate how the scales of the banana cells depend on the Reynolds
numbers, we calculate the power spectra of
the kinetic energy density of the radial flow,
${E}^r_{\rm kin}$,
near the surface ($r=r_s\equiv0.98\, R$). For this, we cut out a thin
latitudinal band around the equator ($\pm\, 7.5\degr$) and calculate the
 power spectrum of ${E}^r_{\rm kin}$, averaged over latitude and time, as a
function of the angular order $m$ for each run.
We rely on the definition
\begin{align}
    \sum_m \widetilde{E}^r_{\rm kin}(m) &= \half\sum_m
  \left\langle{\left|\,{\rm FFT}\left[\big(u_r\!\sqrt{\rho}\big)(r_s)\right]\!(m)\right|^2}\right\rangle_{\theta t} \\
    &= {r_s\over 2}\left\langle{\left(u_r^2\rho\right)(r_s)} \right\rangle_{\theta\phi t} , \nonumber
\end{align}
where the tilde and the operator ${\rm FFT}$ indicate the Fourier transform.
The spectra are shown in \Fig{ban_spec} for all runs, including
the $\Rey$ dependence of the $m$ of their maxima.
Only a weak $\Rey$ dependence is visible.
The spectra of the M runs, except M0, peak around $m=40$, whereas those of the H runs peak
at slightly lower $m\approx32$.
The S run spectra peak at $m=32$ for moderate $\Rm$ and at $m=40$ for high $\Rm$.

\end{appendix}

\end{document}